\documentclass[%
 reprint,
superscriptaddress,
 amsmath,amssymb,
 aps,twocolumn,
 prl,
]{revtex4-2}

\usepackage{graphicx}
\usepackage{dcolumn}
\usepackage{bm}
\usepackage[hypertexnames=false]{hyperref}
\usepackage{siunitx}
\usepackage[capitalise]{cleveref}
\usepackage{lipsum}
\usepackage{flushend}
\usepackage{amsmath}
\usepackage{amsfonts}
\usepackage{amsbsy}
\usepackage{lipsum}
\usepackage{braket}
\usepackage{xcolor}
\usepackage{hyperref}
\usepackage{xr-hyper}
\usepackage{multirow}

\usepackage{braket}

\newcommand{\ah}{\hat{a}}
\newcommand{\hc}{\mathrm{h.c.}}

\usepackage{physics} 
\usepackage{enumerate}
\usepackage{dsfont}

\newcommand{\eff}{\mathrm{eff}}
\newcommand{\opt}{\mathrm{opt}}

\newcommand{\Dicke}[2]{\mathcal{D}^{#1}_{#2}}
\renewcommand{\H}{\mathcal{H}}
\usepackage{xcolor}
\usepackage[normalem]{ulem}

\newcommand{\expect}[1]{\langle {#1} \rangle}
\newcommand{\anticomm}[2]{\{ {#1}, {#2}\}}

\begin{document}


\title{Entanglement-enhanced quantum sensing via optimal global control with neutral atoms in a cavity} 



\author{Vineesha Srivastava}
\affiliation{University of Strasbourg and CNRS, CESQ and ISIS (UMR 7006), aQCess, 67000 Strasbourg, France}

\author{Sven Jandura}
\affiliation{University of Strasbourg and CNRS, CESQ and ISIS (UMR 7006), aQCess, 67000 Strasbourg, France}

\author{Gavin K Brennen}
\affiliation{Center for Engineered Quantum Systems, School of Mathematical and Physical Sciences, Macquarie University, 2109 NSW, Australia}

\author{Guido Pupillo}
\affiliation{University of Strasbourg and CNRS, CESQ and ISIS (UMR 7006), aQCess, 67000 Strasbourg, France}


\date{\today}

\begin{abstract}
We present a deterministic protocol for the preparation of entangled states in the symmetric Dicke subspace of $N$ spins coupled to a common cavity mode that prepares entangled states useful for quantum sensing, achieving a precision significantly better than the standard quantum limit in the presence of photon cavity loss, spontaneous emission and dephasing. The protocol combines a new geometric phase gate which can be utilized for exact unitary synthesis on the Dicke subspace, an analytic solution of the noisy quantum channel dynamics and optimal control methods. This work opens the way to entanglement-enhanced sensing with cold trapped atoms in cavities and is extendable to other spin systems coupled to a bosonic mode.
\end{abstract}
\maketitle
Multi-particle entanglement is an essential resource for achieving quantum advantage in sensing \cite{degen_quantum_2017, pezze2018quantum}, enabling measurement precision variance of the field strength of a signal acting on $N$ spins to be improved from $1/N$ scaling in the standard quantum limit (SQL) to $1/N^2$ scaling in the Heisenberg limit. However, typically the entangled probe states are fragile to errors, posing challenges to quantum sensors that need to be simultaneously sensitive to the unknown field strength they are measuring but insensitive to noise. In the common scenario where the signal overlaps with the noise, even quantum error correction cannot provide an improvement beyond the SQL \cite{Zhou2018,DCS2018} (although see \cite{ouyang2024}).  
Indeed, experiments have so far relied on preparing simpler, spin squeezed states that are somewhat robust to noise~\cite{PhysRevLett.104.073602, schleier2010states, Hosten2016, cox2016deterministic, bohnet2016quantum, pedrozo2020entanglement}. In the ideal scenario without losses, the variance in measurement precision obtained with squeezing from one-axis twisting and two-axis twisting Hamiltonians scale as $1/N^{5/3}$ and $1/N^2$ respectively~\cite{ma_quantum_2011}. While in principle the two-axis twisting generated squeezing in the ideal case can saturate the Heisenberg limit, the amount of achievable squeezing in the experiments is limited by the strength of the interaction Hamiltonian and decoherence~\cite{kitagawa_squeezed_1993}.

In this work we present a simple, deterministic protocol, i.e. not requiring measurement and feedback, to prepare entangled states in the symmetric Dicke subspace of medium sized spin systems $N$ up to $100$ that provide a quantum advantage for sensing and are optimally robust in the presence of noise. 
We focus on spins coupled to a common cavity mode in the regime of strong coupling of cavity quantum electrodynamics, as can be realized, for example, with cold atoms trapped in optical cavities. No direct interactions are required between the spins, though that can provide another handle for Dicke state control \cite{, keating2016arbitrary}. 
Our noise-informed protocol combines a cavity driven {\it geometric phase gate} presented in the companion work~\cite{PhysRevA.110.062610}, 
with an analytical approach to the solution of noisy gate dynamics and optimal control methods to shape the laser pulses -- i.e. the classical photon field driving the cavity mode and the global laser driving collective spin rotations. When applied to the measurement of the strength of an external field, the protocol 
prepares multi-particle entangled states leading to a measurement precision with  significantly better scaling in $N$ than the SQL in the presence of relevant noise, such as photon cavity loss,
spontaneous emission and dephasing already for moderately large strengths of light-matter interactions. 
Surprisingly, the protocol requires only a few global pulses of the cavity mode drive and global rotations, whose parameters we provide. 
We discuss the performance of different classes of entangled states that can be prepared using the protocol for field signal acquisition in the presence of spin dephasing. Using realistic estimates for parameters from current experiments, we find that neutral atoms are excellent candidates for entanglement-enhanced metrology. The approach can be extended to other platforms, e.g. trapped ions or Rydberg atoms.

\begin{figure*}
    \centering
    \includegraphics[width= \linewidth]{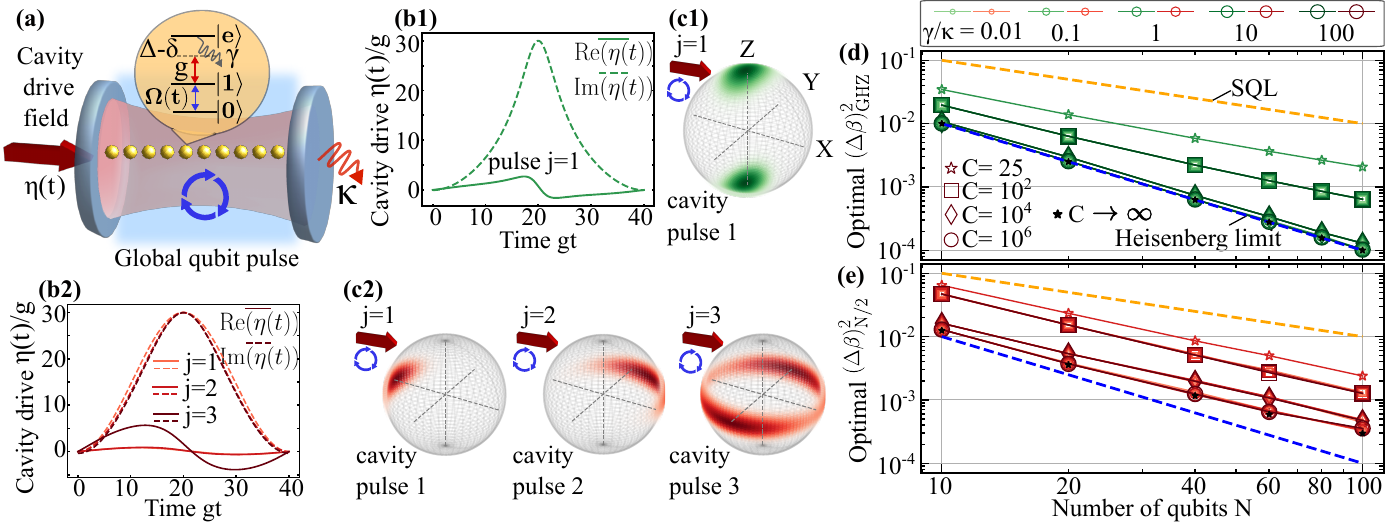}
    \caption{(a) A register of spins with states $\{|0\rangle, |1\rangle, |e\rangle\}$ is coupled to a cavity mode with coupling strength $g$ addressing the $\ket{1}\leftrightarrow \ket{e}$ transition, with detuning $\Delta- \delta$. The cavity mode is externally driven by a laser with amplitude $|\eta(t)|$, and a global laser pulse is applied on  the $\ket{0} \leftrightarrow \ket{1}$ spin transition. Panels (b1,b2): Cavity drive pulses of the optimal state preparation protocol for $N= 40$, $C= 10^4$ and $\gamma/\kappa= 0.01$, for GHZ-like and Dicke-like states, respectively.  Throughout, we make a choice of the cavity drive pulse $\zeta(t)$ in the effective frame with $\mathrm{Re}(\zeta(t))= - 2 \delta \sqrt{\frac{2\phi}{3 \delta T}} \sin^{2}(\frac{\pi t}{T})$ and 
    $\mathrm{Im(\zeta(t))}= - \partial_{t}\mathrm{Re}(\zeta(t))/\delta$ (see \cite{inverting_zeta_note} and \cite{PhysRevA.110.062610}). The obtained minimal measurement precision variances here are $N (\Delta \beta)^2_{\mathrm{GHZ}} = 0.03$ and $N (\Delta \beta)^2_{N/2} = 0.08$. The parameters used in optimal state preparation protocol are listed in the Supplemental Material \cite{a_suppM_note}.
    (c1, c2): State trajectories in Husimi-Q representation of the spin states in the symmetric Dicke subspace after the application of each protocol step $j\, \forall j= 1, \dots, P$. (d) Optimal $(\Delta \beta)^2_{\mathrm{GHZ}}$ for  $P=1$ and (e) $(\Delta \beta)^2_{N/2}$ for $P=3$ obtained as a function of number of qubits $N$, plotted for spin-cavity cooperativities $C= 25$ with $\gamma/\kappa= 1$, and $C= 10^2, 10^4, 10^6$ with different ratios $\gamma/\kappa = 0.01, 0.1, 1, 10, 100$, obtained for the case of $gT \rightarrow \infty$. The optimal states prepared in the presence of finite $C$ successfully surpass the SQL for single particle cooperativity values as small as $C=25$, which has already been achieved in Ref.~\cite{grinkemeyerErrordetectedQuantumOperations2025}.}
    \label{fig:schematic_pulses_results}
\end{figure*}

We consider a setup consisting of $N$ three-level spin systems with computational \textit{qubit} basis states $\ket{0}$ and $\ket{1}$ and an excited state $\ket{e}$. The levels $|1\rangle$ and $|e\rangle$ are coupled via a cavity mode with annihilation (creation) operators $\hat a$($\hat a^{\dagger})$ with coupling strength $g$ (Fig. \ref{fig:schematic_pulses_results}(a)). The cavity mode is driven by a complex classical field of strength $\eta(t)$ which is detuned from the cavity and the $\ket{1}\leftrightarrow\ket{e}$ transition by $\delta$ and $\Delta$, respectively. 
The Hamiltonian reads ($\hbar = 1$)
\begin{eqnarray}\label{eq:H_full}
    &&\hat H = \delta \hat a^{\dagger} \hat a + \left(\Delta - i\frac{\gamma}{2}\right)\hat n_e + \left[ \left(g\hat S^{-} + i\eta(t)\right) \hat a^{\dagger} +\hc\right] \nonumber
\end{eqnarray}
with $\hat n_e = \sum_{j} |e_j\rangle \langle e_j|$, $\hat S^{+}= \sum_j |e_j\rangle \langle 1_j|$, $\hat S^{-}= (\hat S^{+})^{\dagger}$, and $\gamma$ the spontaneous emission rate from $\ket{e}$ state. We account for the decay of the cavity mode at a rate $\kappa$, with the system dynamics governed by the Lindblad master equation (described below). This introduces the single-particle cavity cooperativity, defined as $C = g^2/(\gamma \kappa)$. This approach with spontaneous emission treated as a non-Hermitian contribution gives an upper bound on the obtainable measurement precision~\cite{gamma_note}. 

In the companion work~\cite{PhysRevA.110.062610}, we show that in the
limit of strong cavity driving $\eta/g \rightarrow \infty$ and large detuning $\Delta/g\rightarrow \infty$, and $\delta = \mathcal{O}(g) $, the system dynamics can be reduced to the effective Hamiltonian 
\begin{equation}
    \hat H_{\mathrm{\eff}} = \delta \ah^\dag \ah +  \left(-i\frac{\gamma_1}{2} + \zeta \ah^\dag + \zeta^* \ah\right)\hat{n}_1,
    \label{eq:H_eff}
\end{equation}
with $\hat{n}_1 = \sum_j \ket{1_j}\bra{1_j}$, $ \gamma_{1} = \gamma(1-\sqrt{1-4|\zeta|^2/g^2})/2$, and $\zeta = g^2\alpha/\sqrt{4g^2|\alpha|^2 + \Delta^2}$ where $\dot{\alpha} = -\eta-(i\delta+\kappa/2)\alpha$ with $ \alpha(t=0)=0$. Effective Hamiltonians of a similar form have been previously derived under the condition $g\alpha \ll \Delta$~\cite{PhysRevLett.104.073602, PhysRevLett.116.053601}, whereas our approach works in the wider regime of a large cavity drive and detuning: $ g\alpha \lesssim \Delta$, provided $g \dot \alpha \ll \Delta^2$~\cite{timescales_note}. Interestingly, Eq.~\eqref{eq:H_eff} is equivalent  up to single spin rotations to the M{\o}lmer-S{\o}rensen Hamiltonian~\cite{molmer_multiparticle_1999}, originally developed for trapped ions, and can be thus used to generate fast geometric phase gates -- albeit now for spin systems coupled to a cavity~\cite{PhysRevA.110.062610}. 

We define the quantum channel of the geometric phase gate (realized with a single cavity drive of duration $T$) acting on a basis state $|q_n\rangle \langle q_m|$ of the qubit density matrix, where $\hat n_1 |q_n\rangle = n|q_n\rangle$, ($q_n \in \{0,1\}^N$) as
\begin{equation}
\label{eq:eff_channel}
    \mathcal{E}_{\mathrm{gpg}}(\ket{q_n}\bra{q_m})= e^{i\varphi_{nm}(T)}\ket{q_n}\bra{q_m}.
\end{equation} 
The channel $\mathcal{E}_{\mathrm{gpg}}$ is obtained by considering the open system dynamics determined by Eq.~\eqref{eq:H_eff} 
within a Lindblad master equation approach with $\dot \rho = -i \hat H_{\eff} \rho + i \rho \hat H_{\eff}^{\dagger} + \hat L \rho \hat L^{\dagger} - \{\hat L^{\dagger}\hat L, \rho\}/2$, with $\rho$ the system density matrix and $\hat L = \sqrt{\kappa} \hat a$ the jump operator. The 
cavity is traced out from the joint spin-cavity state, which results in phase accumulation as a function of $n,m$ (i.e., the number of qubits in the $|1\rangle$ state). 
We then combine the dynamics obtained from Eq.~\eqref{eq:eff_channel} with optimal control methods to steer the collective symmetric (Dicke) states of $N$ spin qubits into entangled states of metrological use that are robust to relevant noise sources, such as loss of photons from the cavity mode with rate $\kappa$, loss of population from the excited state $|e\rangle$ with rate $\gamma$ and dephasing in the qubit subspace with rate $\gamma_\phi$. 
This approach consists of the following key steps: (i) Analytically solving the Lindblad master equation for obtaining the geometric phases $\varphi_{nm}(T)$. (ii) Introducing the symmetric Dicke subspace and mapping the geometric phase gate dynamics onto this subspace. (iii) Introducing a state preparation protocol consisting of sequence of pulses where the geometric phase gate operations $\mathcal{E}_{\mathrm{gpg}}$ are combined with global single-qubit rotations to consecutively steer and squeeze an initial coherent Dicke state for a finite number of steps $P$. (iv) Optimizing the state preparation protocol to prepare an arbitrary final state in the symmetric Dicke subspace for a cost function corresponding to the variance of a desired measurement with an observable $\hat M$. The final state acts as the probe state for the defined sensing experiment. 

Depending on the chosen observable $\hat M$, the noise-informed protocol given above leads to the realization of different classes of metrologically useful entangled many-particle states that closely approximate the Heisenberg scaling for realistic values of relevant noise sources, in just one or a few steps $P$. In the following, we demonstrate this for two experimentally relevant choices of $\hat M$. 

{\it The geometric phases $\varphi_{nm}(T)$} [point (i) above] can be obtained analytically  by assuming  $\rho(0) = \ket{0}\bra{0} \otimes \ket{q_{n}}\bra{q_m}$ for the joint cavity-qubit system at time $t=0$. The Lindblad master equation is then exactly solved 
by using the following Ansatz for a state component~\cite{PhysRevA.110.062610} $\rho_{nm}(t) = e^{i\varphi_{nm}(t)} \ket{\beta_n}\bra{\beta_m} \otimes \ket{q_{n}}\bra{q_{m}}/\braket{\beta_n}{\beta_m}$, where $|\beta_n\rangle \langle \beta_m |$ denotes the state of the cavity. Substituting the expression for $\rho_{nm}(t)$ in the Lindblad  master equation, we obtain the following differential equations for $\beta_n(t)$ and $\varphi_{nm}(t)$
\begin{eqnarray}
    \dot{\beta}_n &=& -(i\delta + \kappa/2)\beta_n -in\zeta \label{eq:beta_decay} \\
    \dot{\varphi}_{nm} &=& (m-n)(\zeta\beta_m+\zeta^*\beta_n) +i(m+n)\gamma_1/2 \label{eq:phi_decay}.
\end{eqnarray}
An analytic solution to Eqs.~\eqref{eq:beta_decay} and~\eqref{eq:phi_decay}  is then obtained via an adiabatic approximation in the limit $gT\rightarrow\infty$ and to the first order in $\kappa, \gamma$, by setting $\dot{\beta}_n = 0$ in Eq.~\eqref{eq:beta_decay} as
\begin{eqnarray}
    \label{eq:phase_T_infty}
    \frac{\varphi_{nm}(T)}{\phi} = n^2-m^2 + (m-n)^2\frac{i\kappa}{2\delta} + (m+n)\frac{i\gamma\delta}{2g^2}, 
\end{eqnarray}
where $\phi= \delta^{-1} \int_{0}^{T} \mathrm{d}t |\zeta(t)|^2$ is the \textit{geometric phase} corresponding to the unitary evolution $\hat U_{\mathrm{gpg}} = e^{i\phi \hat n_{1}^2}$ in the lossless case ($\kappa, \gamma= 0$) (the general solution for $\varphi_{nm}(T)$ is given in Supplemental Material~\cite{a_suppM_note}, see also~\cite{PhysRevA.110.062610}).
To our knowledge, this is the first analytic solution of geometric gate dynamics in the presence of relevant noise. 

{\it The Dicke subspace} [point (ii)] is the vector space spanned by states 
$
    |\mathcal{D}^{N}_{n}\rangle = \frac{1}{\sqrt{\binom{N}{n}}}
    \sum_{\{\mathcal{P}~|~\hat n_1 \ket{q_n}= n \ket{q_n}\}} \mathcal{P}|q_n \rangle 
$,
where $\mathcal{P}$ denotes all qubit  permutations resulting in  computational states $|q_n\rangle$ with a fixed number of spins $n$ in $|1\rangle$. 
We note that for a choice of initial state $\ket{\mathcal{D}_0^{N}}$ in the symmetric Dicke subspace, the qubit dynamics during a geometric phase gate remains restricted to the symmetric Dicke subspace. 
The action of the quantum channel $\mathcal{E}_{\mathrm{gpg}}$ on $\rho$ expanded in the Dicke basis then reads 
     $\mathcal{E}_{\mathrm{gpg}}(\rho)=  \sum_{n, m} e^{i \varphi_{nm}(T)} \bra{\Dicke{N}{n}} \rho\ket{\Dicke{N}{m}}\ket{\Dicke{N}{n}}\bra{\Dicke{N}{m}}$, see Eq.~\eqref{eq:eff_channel}.

{\it The state-preparation protocol} [point (iii)] for obtaining  arbitrary $N$-particle entangled states within the Dicke subspace is now realized by a pulse sequence with $P$ steps, where each step $j$ consists of the geometric phase gate $(\mathcal{E}_{\mathrm{gpg}})_j$ followed by a global qubit rotation $\hat{U}_{j}= e^{-i \theta^{\alpha}_j \hat J_z} e^{- i \theta^{\beta}_{j} \hat J_y} e^{-i \theta^{\gamma}_j \hat J_z}$ denoted by map $\mathcal{U}_{j}$, where $\hat J_{\alpha}$ are the collective spin operators in the qubit subspace. 
The corresponding quantum channel $\mathcal{E}_{\mathrm{q}}$ then reads $ \mathcal{E}_{\mathrm{q}}= \mathcal{E}_{P}\cdot \mathcal{E}_{P-1} \cdots \mathcal{E}_{1} \cdot \hat {\mathcal{U}}_0$, with $\mathcal{E}_{j} = \hat{\mathcal{U}}_{j} \cdot (\mathcal{E}_{\mathrm{gpg}})_{j}$. In the limit $T \rightarrow \infty$, $(\mathcal{E}_{\mathrm{gpg}})_{j}$ is fully characterized by the geometric phase $\phi_{j}$ and cavity-drive detuning $\delta_{j}$  for fixed loss rates $\kappa,\gamma$ (see Eq.~\eqref{eq:phase_T_infty}). The state-preparation protocol is thus characterized by the set of parameters $\Theta = \{ \theta^{\alpha}_0, \theta^{\beta}_0, \theta^{\gamma}_0, \theta^{\alpha}_j, \theta^{\beta}_j, \theta^{\gamma}_j, \phi_j, \delta_j \dots; \forall j= 1, 2\dots P \} $, consisting of the global rotation angles $\theta_{j}^{\alpha, \beta, \gamma}$, the geometric phases $\phi_{j}$s and corresponding $\delta_{j}$s in   $\mathcal{E}_{q}$; and is based on optimizing $\Theta$ for a given \textit{protocol cost function}. This approach is similar to recent Refs.~\cite{Johnsson_gpg_2020,gutman2024universal, bond2023efficient}, however, here we find optimal solutions for open quantum systems using noise-informed optimization. Alternative approaches use repeated rounds of active feedback optimization to handle control imperfections~\cite{Marciniak2022, PhysRevX.11.041045}.

In the following, we employ the state-preparation protocol described above to prepare an \textit{optimally robust} probe state for a defined proof-of-principle field-sensing experiment. We define the latter by considering a field along the direction $\Vec{n}$ that is coupled to the $N$ spin qubits  with Hamiltonian $\hat H_{\Vec{n}} = J \hat J_{\Vec{n}}$, with $J$ the coupling strength. $\hat H_{\Vec{n}}$ is applied for a time $t$ such that a given probe state $\rho$ is rotated along the field axis by an angle $\beta= Jt$. 
The goal of the field-sensing experiment is to estimate the rotation angle $\beta$ as accurately as possible by performing measurements on the spins  using an observable $\hat M$. For any given $\hat M$(unbiased estimator), $\beta$ can be estimated with a variance 
\begin{equation}
\label{eq:delta_beta_sq}
    (\Delta \beta)^2 = (\Delta \hat M (\beta))^2/\left| \partial_{\beta} \langle \hat M(\beta)  \rangle \right|^2, 
\end{equation}
where $\hat M(\beta) = e^{-i\hat H_{\Vec{n}} \beta } \hat M e^{i \hat H_{\Vec{n}}\beta}$ and $(\Delta X)^2= \langle X ^2\rangle- \langle X \rangle ^2$. The minimal $(\Delta \beta)^2$ is bound by the quantum Cramer-Rao inequality $(\Delta \beta)^2 \ge 1/\mathcal{F}_{Q}(\rho , \hat H_{\Vec{n}})$,  
where $\mathcal{F}_{Q}(\rho , \hat H_{\Vec{n}})$ is the quantum Fisher information, with $\mathcal{F}_{Q} = N$ and $\mathcal{F}_{Q} = N^2$ for uncorrelated and maximally entangled $N$-spin states, respectively \cite{braunstein1994statistical}. 

The problem we focus on is finding the {\it optimal} probe state that can be prepared in the presence of noise for given $\hat H_{\Vec{n}}$ and $\hat M$ accessible in experiments. This is achieved by choosing $(\Delta \beta)^2$ in Eq.~\eqref{eq:delta_beta_sq} as the \textit{protocol cost function} and minimizing it with respect to $\Theta$ in $\mathcal{E}_q$, for the chosen $\hat M$ [point (iv)], keeping $\beta$ as an additional free parameter in the optimization \cite{extra_rotation_note}. 
 The latter is performed numerically using the \textit{Broyden–Fletcher–Goldfarb–Shanno} method \cite{scheithauerjorge, jones_scipy_2001}, where gradients of the cost function are computed analytically~\cite{a_suppM_note}. Since the optimal parameters $\Theta_{\opt}$ are found for $\kappa, \gamma \neq 0$, the obtained cavity drive and  global qubit pulses are \textit{noise-informed}.

We illustrate the protocol by choosing two different observables $\hat M$ of experimental relevance: (I)  parity along the $x$ axis  $\hat M = \bigotimes_{i=1}^{N}\hat \sigma_{x}^{(i)}$ \cite{leibfried2004toward, schleier2010states}, and (II)  square of the collective spin observable $\hat M= \hat J _z ^2$ along $\hat z$ \cite{lucke2011twin}. Choices (I) and (II) correspond to the  observables that for $\kappa =\gamma = 0$ are theoretically known to saturate the quantum Cramer-Rao inequality with ideal GHZ and Dicke $\ket{\Dicke{N}{N/2}}$ probe states for fields along $\Vec{n}=\hat z$ and $\hat y$ directions, respectively~\cite{toth_quantum_2014, apellaniz2015detecting}. 
\begin{figure}[t!]
    \centering
    \includegraphics[width=\linewidth]{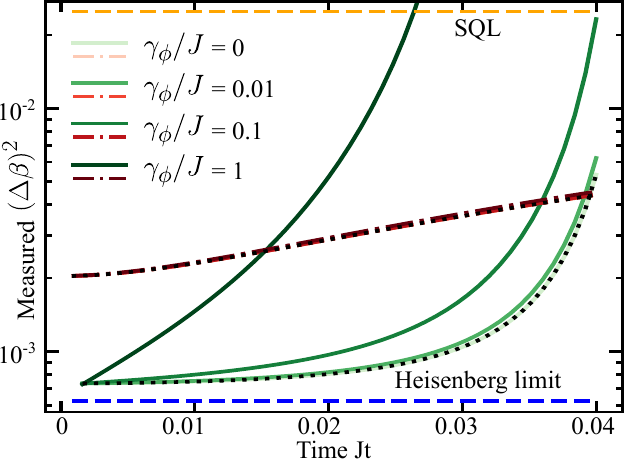}
    \caption{Measured $(\Delta \beta)^2$ as a function of dimensionless signal acquisition time $Jt$ by evolving the optimal probe states under a field coupled with the spins with coupling strength $J$ with local homogeneous dephasing acting on the spins with rates $\gamma_{\phi}/J= 0, 0.01, 0.1, 1.0$ for $N=40$, $C= 10^4$, $\gamma/\kappa = 1.0$. Green solid lines (darker shade for larger $\gamma_{\phi}$) correspond to GHZ-like states while red dash-dot lines correspond to the $\ket{\Dicke{N}{N/2}}$-like states. Dotted black curves are the optimal $(\Delta \beta)^2$ obtained with analytic solution of $\mathcal{E}_{\mathrm{gpg}}$ for $\gamma_{\phi}/J = 0$.} 
    \label{fig:signal_collection}
\end{figure}

We perform extensive numerical simulations in the parameter ranges $10 \leq N \leq 100$, single-particle cooperativities in the range $25 \leq C \leq 10^6$, $10^{-2} \leq \gamma/\kappa \leq 10^2$ for cavity pulse durations $10 \leq gT \leq 10^2$. For both cases (I) and (II), we find that the noise-informed protocol prepares final probe states resulting in measurement variances $(\Delta\beta)^2$ that scale better with $N$ than the SQL in all cases with $C\gtrsim 20$ and closely approach the Heisenberg scaling $(\Delta\beta)^2\sim N^{-2}$ for $C \gtrsim 10^3$, independently of the ratio $\gamma/\kappa$. Pulse durations $gT \lesssim 40$ are sufficient to converge to analytic  results obtained in the adiabatic limit $gT\rightarrow \infty$ from Eqs.~\eqref{eq:eff_channel} and~\eqref{eq:phase_T_infty}, in all shown cases. For each $N$ and $\gamma/\kappa$, $(\Delta\beta)^2$ decreases monotonically with increasing $P$, reaching the analytic predictions for $C \rightarrow \infty$ in just a few steps~\cite{FigS1_note}. The resulting global control pulses have a  smooth, continuous form for all protocol steps $P$. 

Figure~\ref{fig:schematic_pulses_results}(b1) and (b2) show example results of optimal cavity drive pulses $\eta(t)$ found to minimize $(\Delta \beta)^2$ for observables (I) and (II), respectively, for $N=40$, $C= 10^4$, $\gamma/\kappa=0.01$ and $gT= 40$. The plots show a continuous, smooth profile for both real and imaginary parts of $\eta(t)$. 
The protocol for case (I) requires only  $P=1$ step, identically to the noiseless case \cite{monz201114}.  Surprisingly, for case (II), the protocol converges to the asymptotic results in just $P=3$ steps, instead of the generically expected $P\sim N^4$ 
\cite{gutman2024universal} for constructive unitary synthesis or $P\sim N$ \cite{Johnsson_gpg_2020, bond2023efficient} for state synthesis by search.
For each $P$, panels (c1) and (c2) show the corresponding state trajectories in Husimi-Q representation of qubit state in the symmetric Dicke subspace. 
As expected, they appear similar, but not identical, to those of GHZ and symmetric Dicke states: asymmetries due to squeezing-like behavior are visible, resulting on only $\sim 57 \%$ and $\sim 15 \%$ overlap with ideal GHZ and symmetric Dicke states, respectively. Nevertheless, we term them as GHZ-like and $\ket{\Dicke{N}{N/2}}$-like states.
Panels (d) and (e) summarize our results for 
optimal $(\Delta \beta)^2$ as a function of qubit number $N$, for different cooperativities $C$ and  linewidth ratios $\gamma/\kappa$, computed in the limit $gT \rightarrow \infty$. For each $N$, $C$ and $\gamma/\kappa$, the optimization is performed $\mathcal{O}(N)$ times with randomly initialized parameters and the best value is plotted. For case (I) [panel (d)], the optimal probe states prepared with the noise-informed protocol surpass the SQL with variance $(\Delta\beta)^2_{\mathrm{GHZ}}$ scaling with $N$ as $\sim N^{-1.24}$ for cooperativities as small as $C=25$, as $\sim N^{-1.52}$ for $C= 100$, and closely approaching the Heisenberg limit for $C\gtrsim 10^4$, with scaling $(\Delta \beta)^2_{\mathrm{GHZ}} \sim N^{-\alpha}$ and $\alpha > 1.93$. For case (II) [panel (e)], the optimal  $(\Delta \beta)^2_{N/2} \sim  N^{-\alpha}$ scale with $\alpha \approx 1.4$ for $C=25$, $\alpha \approx 1.5$ for $C= 10^2, 10^4$ and $\alpha \approx 1.6$ for $C= 10^6$, showing considerable improvement over the SQL for all $C$. In all cases, optimal results are essentially independent of the ratio $\gamma/\kappa$. 

In order to explore the experimental observability of the above predictions, in Fig.~\ref{fig:signal_collection} we show the performance of the prepared optimal probe states during signal collection in a field-sensing experiment with the field generator $\hat H_{\Vec{n}}$ where spin qubits are additionally subjected to local dephasing with rate $\gamma_{\phi}$, as originated for example by optical trapping of atoms in independent tweezers~\cite{kuhr_analysis_2005}.  
We evolve the optimal probe states prepared for $N=40$, $C= 10^4$ and $\gamma/\kappa= 1.0$ as initial state at $Jt=0$ under the field with the local homogeneous dephasing acting on the spins described as a \textit{collective} process (see Supplemental Material~\cite{a_suppM_note} and Refs.~\cite{PhysRevA.78.052101, shammah2018open}). Figure~\ref{fig:signal_collection} shows that $(\Delta \beta)^2_{\mathrm{GHZ}}$ increases rapidly with time $t$ as $\sim e^{N \gamma_{\phi} t}$ for any given $\gamma_{\phi}/J$ using GHZ-like probe states~\cite{c_suppM_dephasing_results_note, shaji2007qubit}. Results for $\ket{\Dicke{N}{N/2}}-$like states appear instead to be essentially independent of $\gamma_{\phi}/J$ for the shown $t$~\cite{c_suppM_dephasing_results_note}.

Consider an implementation with $^{87}$Rb atoms trapped in optical tweezers and coupled to a fiber Fabry-Perot cavity \cite{grinkemeyerErrordetectedQuantumOperations2025, hunger_fiber_2010-1, uphoff_frequency_2015}. We choose qubit states $\ket{0}= \ket{5 ^{2}S_{1/2}, F= 1, m_{F}= 0}$, $\ket{1}= \ket{5 ^{2}S_{1/2}, F= 2, m_{F}= 0}$, and $\ket{e}= \ket{5 ^{2}P_{3/2}, F= 3, m_{F}= 0}$, where the linewidth of the $\ket{1} \leftrightarrow \ket{e}$ transition ($\lambda = 780\,$nm) is $\gamma =2\pi \times 6\,$MHz (FWHM). We assume a cavity finesse $F \approx 2 \times 10^5$, a waist radius ${\rm w}_r \approx 2\,\mu$m and a length $L \approx 40\,\mu$m resulting in a cooperativity of $C = 3\lambda^2F/(2\pi^3 {\rm w}_r^2) \approx 1500$ with a coupling strength of $g = \sqrt{3\lambda^2 C \gamma}/(2\pi^2 {\rm w}_r^2L) \approx 2\pi \times  400\,$MHz and $\kappa =\pi C/L F \approx 2\pi \times 20\,$MHz (FWHM), so that $\gamma/\kappa \approx 0.3$~\cite{TC_validity_note}. Our noise-informed state preparation protocol obtains for $N=10$ atoms a minimal $(\Delta \beta)^2_{N/2} = 0.022$ with $P=3$ protocol steps and a minimal $(\Delta \beta)^2_{\mathrm{GHZ}} = 0.013$ with $P=1$ protocol step, where in each step the cavity pulse is applied for a duration $T= 20 g^{-1} \approx 8\,$ns. Tweezer induced dephasing rates on state $\ket{1}$ can be as small as $\gamma_{\phi}/g = 0.03 \times 10^{-6}$ \cite{manetsch2024tweezer}, which we find to be negligible~\cite{a_suppM_note}. 
Our protocol is also applicable to platforms utilizing circular Rydberg transitions coupled to a microwave cavity. Recent advances demonstrate a high-finesse microwave resonator predicting a single particle cooperativity of $C = 6.75 \times 10^5$~\cite{zhang2025opticallyaccessiblehighfinessemillimeterwave}. 

To compare our protocol's performance with prior works on spin squeezing with neutral atoms in cavity, we note that the most recent experiment with $\sim 400$ Yb-171 atoms achieves an improvement of $\mathcal{\xi}^2 \approx -4.5$ dB~\cite{pedrozo2020entanglement}. With our protocol to generate the maximally squeezed state on the equator of the collective Bloch sphere, we can theoretically achieve $\xi^2 = N (\Delta \hat J_z^2)/ \langle \hat J_y^2 \rangle^2 \approx -12.72$ dB with $N=100$ atoms for a single particle cooperativity of $C=25$, which has already been demonstrated~\cite{grinkemeyerErrordetectedQuantumOperations2025}.
Moreover, the only demonstration surpassing the SQL with neutral atoms via genuine multi-partite entanglement was achieved without cavities, using a GHZ-state generation method that is not scalable beyond ten atoms~\cite{caoMultiqubitGatesSchrodinger2024b}.

Finally, the setup described above is sufficient to achieve unitary synthesis in the Dicke subspace. The control algebra $\{\hat J_z^2,\hat J_x,\hat J_y,\hat J_z\}$ is universal for Dicke state preparation starting from a canonical product state like $\ket{D^N_N}$ \cite{phdthesis_Merkel_2009} and is efficient \cite{gutman2024universal, bond2023efficient}. By a simple modification \cite{protocol_B_note} that enables multi-controlled phase gates, our protocol is exactly  universal for such unitary synthesis. 

\begin{acknowledgments}
This research has received funding from the European Union’s Horizon 2020 research and innovation programme under the Marie Sklodowska-Curie project 847471(QUSTEC) and project 955479 (MOQS), the Horizon Europe programme HORIZON-CL4-2021-DIGITAL-EMERGING-01-30 via the project 101070144 (EuRyQa) and from the French National Research Agency under the Investments of the Future Program
projects ANR-21-ESRE-0032 (aQCess), ANR-22-CE47-0013-02 (CLIMAQS) and QuanTEdu-France. G.K.B. acknowledges support from the Australian Research Council Centre of Excellence for Engineered Quantum Systems (Grant No. CE 170100009). Computing time was provided by the High-Performance Computing Center of the University of Strasbourg.
\end{acknowledgments}




\bibliography{library_arxiv_v3}

\begin{thebibliography}{53}
\bibitem [{inv()}]{inverting_zeta_note}%
  \BibitemOpen
  \href@noop {} {}\bibinfo {note} {The cavity drive pulse $\eta(t)$ in the original frame is obtained by inverting the pulse $\zeta(t)$ in Eq.~\eqref{eq:H_eff} with a finite value of $\Delta \gg g$ which is set by a choice of $\mathrm{max}_{t}|\eta(t)| \gg g$: $\eta(t)=-\frac{\Delta}{S(t)}(\dot{\zeta}(t)+(i\delta+\frac{\kappa}{2})\zeta(t))-\frac{2\Delta g^2}{S(t)^3}\zeta(t)\frac{d}{dt}|\zeta(t)|^2$, where $S(t)=g^2\sqrt{1-4|\zeta(t)|^2/g^2}$. Here, we make a choice of $\mathrm{max}_{t}(|\eta(t)|)= 30g$.}\BibitemShut {Stop}%
\bibitem [{ext()}]{extra_rotation_note}%
  \BibitemOpen
  \href@noop {} {}\bibinfo {note} {Without loss of generality, we set $\beta_{\opt}=0$ by adding a step in the protocol corresponding to global qubit rotation by the found $\beta_{\opt}$ along the field-axis $\Vec{n}$ (known)}\BibitemShut {NoStop}%
\bibitem [{pro()}]{protocol_B_note}%
  \BibitemOpen
  \href@noop {} {}\bibinfo {note} {In the companion work Ref.~\cite{PhysRevA.110.062610} we propose an alternative adiabatic phase gate using the same cavity setup as above but in the weak drive, long pulse time limit with detunings $\Delta,\delta=O(g)$. There we show that sequential application of $N$ adiabatic gates can generate a multi-controlled phase gate deterministically $V(\phi)=e^{i \phi \ket{D^N_N}\bra{D^N_N}}$. Writing an arbitrary unitary in the Dicke subspace in its spectral decomposition, $U=\sum_{j=1}^{N+1} e^{i \lambda_j \ket{\lambda_j}\bra{\lambda_j}}$, the following decomposition suffices $U=\prod_{j=1}^{N+1} W(\lambda_j)^{\dagger}V(\lambda_j)W(\lambda_j)$, where $W(\lambda_j)^{\dagger}$ is any unitary extension of the state mapping $W(\lambda_j)^{\dagger}\ket{D^N_N}=\ket{\lambda_j}$.}\BibitemShut {Stop}%
\end{thebibliography}
%

\newpage
\pagebreak

\onecolumngrid

\newcommand{\cav}{\mathrm{cav}}
\newcommand{\id}{\mathrm{id}}
\newcommand{\cc}{\mathrm{c.c.}}
\newcommand{\state}[3]{\ket{{#1}_1{#2}_e{#3}_{\mathrm{ph}}}}
\newcommand{\statebra}[3]{\bra{{#1}_1{#2}_e{#3}_{\mathrm{ph}}}}
\newcommand{\f}{E}
\renewcommand{\ne}{\hat{n}_e}
\renewcommand{\Im}{\mathrm{Im}}
\newcommand{\lf}{\left(\frac{\lambda}{2}\right)}
\renewcommand{\H}{\mathcal{H}}
\renewcommand{\ne}{\hat{n}_e}
\renewcommand{\Im}{\mathrm{Im}}

\vfill
\pagebreak

\setcounter{equation}{0}
\setcounter{figure}{0}
\setcounter{table}{0}
\setcounter{section}{0}
\renewcommand{\theequation}{S\arabic{equation}}
\renewcommand{\thefigure}{S\arabic{figure}}

\center{\large{\textbf{Supplemental Material}}}

\raggedright

\section{Summary of main results}

In this section, we summarize the main results of the state-preparation protocol for entanglement-enhanced quantum sensing, which prepares an entangled probe state in the presence of noise. The sensing experiment is defined by a fixed field $\hat{H}$ and a measurement observable $\hat{M}$, which are designed to estimate the parameter $\beta$—the rotation angle caused by the field $\hat{H}$ on $N$ spins prepared in the probe state. We prepare two different classes of optimal entangled probe states using a finite number of cavity drive pulses, ensuring that the prepared state achieves minimal variance in the estimation of $\beta$, characterized by $(\Delta \beta)^2$. The optimal $(\Delta \beta)^2$ scales as $\sim N^{-\alpha}$. The results for different cooperativities $C$ are summarized in TABLE~\ref{tab:summary_table} below; see also Fig.~\ref{fig:schematic_pulses_results}(d) in the main text for reference.
\begin{table}[h!]
    \centering
    \begin{tabular}{|c|c|c|c|c|c|c|c|c|} \hline 
         & \multirow{2}{*}{Field $\hat H$} &  \multirow{2}{*}{Measurement $\hat M$}  & \multirow{2}{*}{Number of cavity pulses} &  \multirow{2}{*}{Probe state}  & \multicolumn{4}{c|}{Optimal $(\Delta \beta)^2 \sim N^{-\alpha}$}\\ \cline{6-9}
         &   &    & &  & $C = 25$ & $C= 10^2$ & $C= 10^4$ & $C= 10^6$\\
         \hline 
        Case I & $J \hat J_z$ & $\bigotimes_{i=1}^{N} \hat \sigma_{x}^{(i)}$  &1& GHZ-like & $\alpha= 1.24$ & 1.52 & 1.93 & 1.99\\ \hline 
        Case II & $J \hat J_y$ & $\hat J_z^2$  &3& $|\mathcal{D}_{N/2}\rangle$-like & $\alpha = 1.4$ & 1.5 & 1.5 & 1.6 \\ \hline
    \end{tabular}
    \caption{Summary of the key results for the state-preparation protocol in entanglement-enhanced quantum sensing. The table compares two cases (Case I and Case II) in terms of the field $\hat{H}$, measurement observable $\hat{M}$, the number of cavity drive pulses, the type of entangled probe state prepared, and the scaling of the optimal variance $(\Delta \beta)^2 \sim N^{-\alpha}$, where the optimizations are performed for up to $N= 100$, and in the limit $gT \rightarrow \infty$ of the cavity drive duration. The scaling parameter $\alpha$ is provided for different cooperativities $C = 25$, $10^2$, $10^4$, and $10^6$, highlighting the entanglement-enhanced advantage of the estimation precision over SQL ($\alpha =1$).}
    \label{tab:summary_table}
\end{table}

\section{Estimates of cavity parameters and validity of Tavis - Cummings Hamiltonian for neutral atoms coupled to optical Fiber Fabry Perot cavity}

In this section, we analyze the parameter regimes where our protocol can be realized in the context of neutral atom spin qubits coupled to an optical cavity mode. We consider here the example of the D2 line in $^{87}$Rb atoms at $\lambda = 780\,$nm with linewidth $\gamma = 2\pi \times 6\,$MHz coupled to an optical Fabry Perot fiber cavity (FPFC), placed with $\approx 5\,\mu$m spacing similar to Ref.~\cite{grinkemeyerErrordetectedQuantumOperations2025}. Hence as the number of spins $N$ increase, one would require a cavity of length $5N\, \mu$m. 

\begin{figure}[h!]
        \centering
        \includegraphics[width=0.5\linewidth]{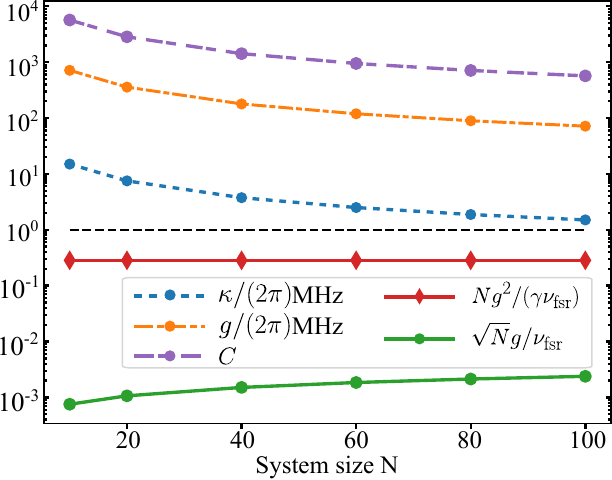}
        \caption{ Estimates of cavity photons loss rate $\kappa$, atom- cavity coupling strength $g$ and single particle cooperativity $C$ as a function of system size $N$. Additionally we plot the ratios $\sqrt{N}g/\nu_{\rm fsr}$ (green circle markers) and $Ng^2/(\gamma \nu_{\rm fsr})$ (red diamond markers) to check the validity of the Tavis Cummings model for moderate sized systems of neutral atom spin qubits with $^{87}$Rb coupled to optical FPFC cavity mode.}
        \label{fig:FPFC_estimates}
    \end{figure}

For a cavity of length $L$, and radii of curvatures at the input and output mirrors $R_1$ and $R_2$, by defining $g_1= 1 - L/R_1$, $g_2 = 1- L/R_2$, the waist spot size is given as~\cite{siegmanSiegmanLasersUniversityScience} 
\begin{equation}
    w_0 = \sqrt{\frac{L\lambda}{\pi}}\left(\frac{g_1g_2(1-g_1g_2)}{(g_1 + g_2 - 2g_1g_2)^2}\right)^{1/4}
\end{equation}
For a stable cavity, one requires also $0\leq g_1g_2 \leq 1$, and taking into account the current fabrication limits corresponding to $R = 50\,\mu$m~\cite{hunger_fiber_2010-1}, we make a choice of $R_1 = R_2 = 2.1(L/2)$ for system sizes $N = 10 -100$. With these considerations and assuming a maximal achievable cavity Finesse of $F = 200\,000$, and ignoring the cavity clipping losses, we plot in Fig.~\ref{fig:FPFC_estimates} the achievable coupling strength $g = \sqrt{\frac{3\lambda^2 c \gamma}{2 \pi^2 w_0^2 L}}$, $\kappa = \nu_{\rm fsr}/F$, $C = g^2/(\kappa \gamma)$ and the two ratios to validate the Tavis Cummings model given by $\sqrt{N}g/\nu_{\rm fsr}$ and $Ng^2/(\gamma \nu_{\rm fsr})$ as a function of the system size $N$, where $\nu_{\rm fsr} = 2\pi c/(2L)$ is the cavity free spectral range. The parameters satisfy the conditions $\sqrt{N}g/\nu_{\rm fsr} \ll 1$ and $Ng^2/(\gamma \nu_{\rm fsr}) \ll 1$, hence validating the Tavis Cummings model~\cite{blahaTavisCummingsModelRevisiting2022}. 

\section{Timescale and strength constraints in deriving \texorpdfstring{$\hat H_{\rm eff}$}{H eff(t)}}

In this section, we review the derivation of the effective Hamiltonian $\hat{H}_{\rm eff}$, presented in Eq.\eqref{eq:H_eff} of the main text, which underpins the implementation of the geometric phase gate. This derivation follows the detailed approach outlined in Ref.\cite{PhysRevA.110.062610}. We also examine the key timescales, energy scales and constraints crucial for implementing the geometric phase gate within the system, with a summary provided at the end of this section.

The derivation begins with the full Hamiltonian, expressed as:
    \begin{equation}
        \hat H = \delta \ah^{\dag} \ah+ \left(\Delta - i\frac{\gamma}{2}\right)\hat n_e + \left[\left(g\hat S^{-} + i \eta(t)\right)\ah^{\dag} + \hc \right].
    \end{equation}
    The system dynamics are described by the Lindblad master equation $\dot{\rho} = -i\hat H\rho +i\rho \hat H^\dagger+ \hat L\rho \hat L^\dagger- \{\hat L^\dagger\hat L, \rho\}/2$ with the jump operator $\hat L=\sqrt{\kappa}\ah$.
    
    Starting from this, the effective Hamiltonian $\hat H_{\rm eff}$ in Eq.~\eqref{eq:H_eff} (in the main text) for the geometric phase gate is derived in the limit of $\Delta/g, \eta/g \rightarrow \infty$ with $\delta$ set constant in the order $\mathcal{O}(g)$. The cavity drive pulse duration $T$ is of the order $\mathcal{O}(g^{-1})$. The effective Hamiltonian is derived in the following two steps:

    \begin{enumerate}
        \item Displace the cavity mode with a time-dependent displacement operator $\hat D(\alpha(t)) = \exp{(\alpha(t) \ah^\dagger- \alpha^*(t) \ah)}$, where $\dot \alpha = -\eta - (i\delta + \kappa/2)\alpha$. The Hamiltonian after this transformation reads 
        \begin{equation}
            \Tilde{\hat H} = \delta \ah^{\dag}\ah + (\Delta - i\gamma/2)\hat n_e + g(\ah^{\dag}\hat S^{-} + \ah \hat S^{+}) - g\alpha^*\hat S^{-} - g\alpha \hat S^{+}.
            \label{eq:sm_Hamiltonian_trasnformed_cavity_frame}
        \end{equation}

        In writing this, there is no approximation or elimination of terms based on timescales. This transformation is introduced to reduce the number of cavity photons, and precisely corresponds to moving into a reference frame in which the cavity remains in the vacuum state when no qubit is coupled to it, i.e. when no spin qubit is in the the state $|1\rangle$.

        \item Diagonalise the spin subspace to derive an effective Hamiltonian that simplifies the system's dynamics by focusing only on the relevant terms, assuming the system starts with no spins in the $|e\rangle$ state. Using the adiabatic theorem, we analyze the timescales governing the evolution. Considering the part of $\Tilde{\hat H}$ that scales with $\Delta$, we have 
            \begin{equation}
                \hat H^{(0)}(t) = \Delta \hat n_e - g\alpha(t) \hat S^{+} - g\alpha(t)^* \hat S^{-}.
                \label{eq:H_0}
            \end{equation}
            Since $\hat H^{(0)}$ does not couple any two spins and commutes with the total excitation number operator $\hat n = \hat n_1 + \hat n_e$, it can be written as a sum of $n$ single spin Hamiltonians. Explicitly,
            \begin{equation}
                \hat H^{(0)}(t) = \sum_{j=1}^{n}\left(\Delta |e_j\rangle\langle e_j| - g\alpha(t) |e_j\rangle \langle 1_j| - g\alpha(t)^* |1_j\rangle \langle e_j|\right).
            \end{equation}

            The lowest and the first excited instantaneous eigenvalues of $\hat H^{(0)}(t)$ are given by:
            \begin{eqnarray}
                E_{0} &=& \frac{n}{2}(\Delta - \sqrt{\Delta^2 + 4g^2|\alpha|^2}),\\
                E_{1} &=& \frac{n-1}{2}(\Delta - \sqrt{\Delta^2 + 4g^2|\alpha|^2}) + \frac{1}{2}(\Delta + \sqrt{\Delta^2 + 4g^2|\alpha|^2}),
            \end{eqnarray}
            yielding an energy gap of:
            \begin{equation}
                E_{\rm gap} = \sqrt{\Delta^2 + 4g^2|\alpha|^2} = \mathcal{O}(\Delta).
            \end{equation}
            
            Restricting our analysis to the local state space orthogonal to $\ket{0}$ everywhere (since the full Hamiltonian acts trivially on that state), the instantaneous eigenstate corresponding to $E_0$ is 
            \begin{equation}
                |\psi_{0}; t\rangle = (\cos(\theta/2)|1\rangle - \sin(\theta/2) e^{-i\varphi}|e\rangle)^{\otimes n},
            \end{equation}
            where $\cos(\theta) = \frac{\Delta}{\sqrt{\Delta^2 + 4g^2|
            \alpha(t)|^2}}$ and $\varphi = {\rm arg}(\alpha(t))$. 

            By choosing a cavity drive such that $|\alpha(0)| = 0$ and $|\alpha(T)| = 0$, the instantaneous eigenstates at $t=0$ and $t=T$ correspond to $|\psi_{0}\rangle = |1\rangle^{\otimes n}$. Thus, starting in the state $|1\rangle^{\otimes n}$ at $t=0$ and adiabatically following $|\psi_{0}; t\rangle$ requires satisfying: 
            \begin{equation}
                 \frac{\langle \psi_{0}; t| \frac{{\rm d} \hat H^{(0)}(t)}{{\rm d}t}| \psi_{1}; t\rangle }{E_{\rm gap}} \ll \langle \psi_0(t)|\left[ \frac{\partial}{\partial t} |\psi_0; t\rangle \right] \sim E_0
            \end{equation}
            This condition is simplified to:
            \begin{equation}
                g \dot \alpha \ll \Delta^2,
            \end{equation}
            and, since $\Delta/g\gg 1$ by assumption, is surely satisfied when 
            \begin{equation}
                \dot \alpha \ll \Delta,
            \end{equation}
            implying, that when $\alpha(t)$ remains nearly constant on timescales much longer than $1/\Delta$, adiabatic evolution is guaranteed.

            We now derive the effective Hamiltonian by projecting the full Hamiltonian $\Tilde{\hat H}$ onto the lowest instantaneous eigenstate $|\psi_{0}; t\rangle$:
            \begin{equation}
                \hat H'_{\rm eff} = \langle \psi _{0}; t| \Tilde{\hat H} |\psi_{0}; t\rangle.
            \end{equation}
            
            We obtain,
            \begin{align}
        \begin{split}
            \hat H'_{\rm eff} &= \delta \ah^{\dag}\ah + \left(\epsilon_1 - i\gamma_1/2\right)\hat n_1 + \left(\epsilon_e - i\gamma_e/2\right)\hat n_e \\
            &+ \frac{g^2}{\sqrt{\Delta^2 + 4g^2|\alpha|^2}} (\alpha \ah^{\dag} + \alpha^*\ah)(\hat n_1 - \hat n_e) + \mathcal{O}(g) \hat S^{+} + \mathcal{O}(g)\hat S^{-},
        \end{split}
        \end{align}
        where $\epsilon_{e/1}= (\Delta \pm \sqrt{\Delta^2 + 4g^2|\alpha|^2})/2 +\mathcal{O}(g)$, $\gamma_{e/1} = \gamma(1\pm \sqrt{1- 4|\zeta|^2/g^2})/2$ and $\mathcal{O}(g)$ refers to the terms which do not diverge in the limit $\Delta/g \rightarrow \infty$. 
        By neglecting terms of order $\mathcal{O}(g)$ that act only on the spins (without involving $\ah$ or $\ah^{\dag}$), and by eliminating terms proportional to $\hat n_e$ under the assumption that the system starts with no spins in the $|e\rangle$ state, the effective Hamiltonian simplifies to 
        \begin{equation}
            \hat H_{\rm eff}= \delta \ah^{\dag}\ah + \left( -i\frac{\gamma_1}{2} + \zeta(t)\ah^{\dag} + \zeta^*(t) \ah\right)\hat n_1,
        \end{equation}
        where we define:
        \begin{equation}
            \zeta(t) = \frac{g^2\alpha(t)}{\sqrt{4g^2|\alpha|^2 + \Delta^2}}.
            \label{eq:zeta_alpha}
        \end{equation}
    \end{enumerate}
    
    Thus, the effective Hamiltonian $\hat H_{\rm eff}$ contains the spin-cavity interaction dynamics while assuming adiabatic evolution, negligible contributions from $\hat n_e$, and elimination of terms scaling as $\mathcal{O}(g)$ acting only on spins. It is worth noting that in Ref.~\cite{PhysRevA.110.062610}, the second step is presented differently, where the diagonalization of the spin subspace is instead described in terms of a basis transformation within the spin subspace.

    \subsubsection*{Estimating number of photons in the transformed cavity frame}
    Let the state of the cavity mode corresponding to $n_1$ spins in $|1\rangle$ at any time $t$ be denoted by $\rho_{n_1}(t)$. It evolves under the Lindblad equation $\dot \rho_{n_1} = - i \left[ \hat H_{n_1} , \rho_{n_1}\right] + \hat L \rho_{n_1} \hat L^{\dagger} - \left\{ \hat L^{\dagger} \hat L, \rho_{n_1}\right\}/2$ with $\hat H_{n_1} = \delta \ah^{\dag} \ah + \left(n_1\zeta(t) \ah^{\dagger} + \hc \right)$ and $\hat L = \sqrt{\kappa}\ah$. Assuming an initial coherent pure state for the cavity mode and spins in the eigenspace $n_1$, the cavity mode remains in a pure state at all times, even if it undergoes decay. Hence we use the Ansatz for $\rho_{n_1}(t)$ given by
    \begin{equation}
    \label{eq:si_cav_state_ansatz}
        \rho_{n_1}(t) = | -\alpha(t)\rangle \langle -\alpha(t)|,
    \end{equation}

    On substituting the ansatz from Eq.~\eqref{eq:si_cav_state_ansatz} in the Lindblad equation, we obtain on the left-hand side

    \begin{equation}
        {\rm LHS}= \dot \rho_{n_1} = -\dot \alpha \ah^{\dagger} \rho_{n_1} - \dot \alpha^* \rho_{n_1}\ah - \frac{d|\alpha|^2}{dt}\rho_{n_1},
    \end{equation}
    where we use the property of a coherent state that $\frac{d}{dt} |\alpha(t)\rangle = \dot \alpha \ah^{\dag} |\alpha\rangle - \frac{1}{2} \frac{d|\alpha|^2}{dt} |\alpha\rangle$. The right-hand side is obtained as 
    \begin{equation}
        {\rm RHS} = ([(i\delta + \kappa/2) \alpha - i n_1 \zeta(t)] \ah^{\dagger} \rho_{n_1} + \hc ) + (in_1 \zeta^*(t) \alpha - i n_1 \zeta (t) \alpha^* + \kappa |\alpha|^2) \rho_{n_1}
    \end{equation}
    On equating the above two equations, we obtain the solution given by 
    \begin{equation}
        \dot \alpha = i n_1 \zeta(t) - (i \delta + \kappa/2)\alpha 
        \label{eq:sm_alpha_solution}
    \end{equation}
    In the steady state, we have $\dot \alpha = 0$, so Eq.~\eqref{eq:sm_alpha_solution} yields an average number of photons of 
    \begin{equation}
        \langle \ah^{\dagger} \ah\rangle = |\alpha|^2 = \frac{|\zeta|^2 n_1^2 }{\delta^2 (1+ \kappa/(4\delta^2))} \approx \frac{|\zeta|^2n_1^2}{\delta^2},
    \end{equation}
    and since we have $\Delta \gtrsim g|\alpha|$, $|\zeta| = \mathcal{O}(g)$, and hence for moderately sized systems, 
    \begin{equation}
        \delta \sqrt{\langle \ah^{\dag} \ah\rangle} = \mathcal{O}(g).
    \end{equation}


    \subsection{Implementation of geometric phase gate with \texorpdfstring{$\hat H_{\rm eff}$}{H eff(t)}}

    To realize the geometric phase gate using $\hat H_{\rm eff}$, we choose $\zeta(t)= -\delta f(t) + i \dot f(t)$, where $f(t) = 2\sqrt{\frac{2\phi}{3\delta T}} \sin^2\left(\pi t/T\right)$.  This choice ensures that the effective Hamiltonian implements the unitary operation $(\kappa, \gamma= 0)$ given by $\hat U_{\rm gpg} = e^{i\phi \hat n_1^2}$, with $\phi = \delta \int_{0}^{T} {\rm d }t f(t)^2$~\cite{PhysRevA.110.062610}. Note that $f(0) = f(T) =0$, and $\dot f(0) = \dot f(T) = 0$, which guarantees $\alpha(0)= \alpha(T)=0$, consistent with the adiabatic theorem required to derive $\hat H_{\rm eff}$ as discussed above. 
    
    For an initial joint state of the cavity-qubit system given by $|\psi(0)\rangle = |\beta(0)\rangle \otimes |q_n\rangle$, where $|\beta(0)\rangle$ represents the initial cavity state and $|q_n\rangle$ is the computational qubit state with $\hat n_1 |q_n\rangle = n |q_n\rangle$, the choice of $\zeta$ ensures that after the application of $\hat U_{\rm gpg}$, the final state becomes $|\psi(T)\rangle= e^{i\phi \hat n_1^2} |\beta(0)e^{-i\delta T}\rangle\otimes |q_n\rangle$~\cite{PhysRevA.110.062610}. Notably, the action of $\hat U_{\rm gpg}$ is independent of the initial cavity state $\beta(0)$. Since any arbitrary initial state of the cavity can be written as a superposition of different coherent states $|\beta(0)\rangle$, the geometric phase gate unitary operation is independent of the initial state of the cavity.  
    

    \subsection{Constraints on the effective cavity drive \texorpdfstring{$\zeta(t)$}{zeta(t)}}\label{subsec:zeta_constraint_delta_bounds}
    
    Rearranging Eq.~\eqref{eq:zeta_alpha}, we can express $\alpha(t)$ in terms of $\zeta(t)$
    \begin{equation}
        |\alpha|= \frac{\Delta |\zeta|}{\sqrt{g^4 - 4g^2|\zeta|^2}},
        \label{eq:alpha_zeta}
    \end{equation}
    which establishes the following constraint to ensure the cavity drive can be inverted back to the laboratory frame:
    \begin{equation}
        |\zeta(t)| < \frac{g}{2}.
        \label{eq:zeta_constraint}
    \end{equation}

\begin{figure}[h!]
        \centering
        \includegraphics[width=0.5\linewidth]{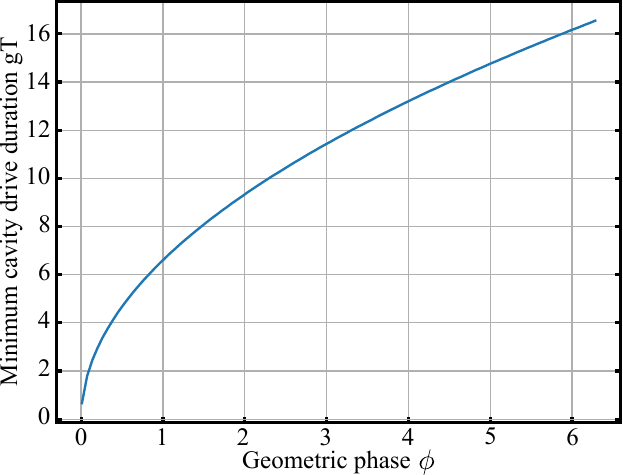}
        \caption{ Minimum $gT$ required to implement a geometric gate $\hat U = e^{i\phi \hat n_1^2}$ plotted as a function of the geometric phase $\phi$ such that a $\delta$ exists in accordance with the inequality $|\zeta(t)|< g/2$, see Eq.~\eqref{eq:delta_range}.}
        \label{fig:min_gT_plot}
    \end{figure}

    From this, we derive a bound for $\delta$ for a given $\phi$ and $T$:
    \begin{equation}
        \delta \in \left(\frac{2\pi}{T}, \frac{3g^2T}{32\phi}\right).
        \label{eq:delta_range}
    \end{equation}
    Figure~\ref{fig:min_gT_plot} illustrates the minimum $gT$ required for a $\delta$ to exist (within the given bounds), depending on the value of $\phi$ implemented by the geometric phase gate. 

We now present a summary of the strength conditions and timescales:
    \begin{itemize}
        \item $\Delta/g \rightarrow \infty$, $\eta/g \rightarrow \infty$.\\
        \item Cavity drive pulse duration $T = \mathcal{O}(g^{-1})$. \\
        \item $\delta = \mathcal{O}(g)$ and $\delta \in \left(\frac{2\pi}{T}, \frac{3g^2T}{32\phi}\right)$ for a geometric phase gate with phase $\phi$. 
        \item $g|\alpha(t)| \lesssim \Delta$ : Rabi frequency of the effective cavity mediated transition on the spins is comparable to the detuning of the transition.\\
        \item $g \dot \alpha \ll \Delta^2$ : evolution timescale to eliminate the excited state subspace with $|e\rangle$.\\
    \end{itemize}
    
\section{Exact solution of the geometric phases in the presence of losses}

In this section, we present the exact solution of the geometric phases $\varphi_{nm}(T)$ in Eq.~\ref{eq:eff_channel} (of the main text).

We describe the state of the joint spin-cavity system at any time $t$ as  $\rho(t)= \sum_{n,m} \rho_{nm}(t)$, and use an Ansatz for the state components $\rho_{nm}(t)$ given by 
\begin{eqnarray}
    \rho_{nm}(t) = e^{i\varphi_{nm}(t)} \frac{|\beta_n(t)\rangle\langle \beta_m(t)| \otimes |q_{n}\rangle \langle q_{m}|}{\langle \beta_m(t) |\beta_n(t) \rangle},
    \label{eq:ansatz_rho}
\end{eqnarray}
where $\varphi_{nm}(t)$ are the \textit{geometric-phases} acquired by the qubit state component $\ket{q_{n}}\bra{q_{m}}$, and $ \ket{\beta_n}\bra{\beta_m}$ is the corresponding state of the cavity mode. With this Ansatz, we exactly solve the open quantum system for $\rho_{nm}(t)$ with $\kappa, \gamma \neq 0$. The latter is described by the Lindbladian master equation given by
\begin{eqnarray}\nonumber
    \dot{\rho}_{nm} &= - i \hat H_{\eff}^{n} \rho_{nm} + i \rho_{nm} (\hat H_{\eff}^{m})^{\dagger} + \hat L \rho_{nm} \hat L^{\dagger} - \frac{1}{2}\{ \hat L^{\dagger} \hat L, \rho_{nm}\}, 
\end{eqnarray}

with $\hat H_{\eff}^{n}= \delta a^{\dag}a + (-i\frac{\gamma_1}{2} + \zeta(t)\ah^{\dag} + \zeta(t)^* \ah) n$. 
On substituting the Ansatz for $\rho_{nm}$ in the master equation, we obtain the derivatives for $\beta_n(t)$ and $\varphi_{nm}(t)$ as (see also Ref. \cite{PhysRevA.110.062610}),
\begin{eqnarray}
    \dot{\beta}_n &=& -(i\delta + \kappa/2)\beta_n -in\zeta \label{eq:beta_decay_sm}, \\
    \dot{\varphi}_{nm} &=& (m-n)(\zeta\beta_m+\zeta^*\beta_n) +i(m+n)\gamma_1/2. \label{eq:phi_decay_sm}
\end{eqnarray}

We now take the initial state of the joint spin-cavity system as $\rho (0) = \ket{\beta_{n}(0)}\bra{\beta_{m}(0)} \otimes \ket{q_n}\bra{q_m}$, which forms the basis for all possible initial states and is hence sufficient to obtain a general solution for the state evolution. The solutions corresponding to $\beta_n(t)$ and $\psi_{nm}(t)$ are then given by
\begin{equation}
    \beta_n(t) = \beta_{n}(0)e^{-(i\delta+\kappa/2)t} -in \int_0^t \mathrm{d}t' \zeta(t') e^{-(i\delta+\kappa/2)(t-t')} 
    \label{eq:beta_soln},
\end{equation}
and
\begin{eqnarray}
    \varphi_{nm}(t) = \int_0^t  &\Big[& (m-n)(\zeta(t) \beta_m(t)^* + \zeta(t)^*\beta_n(t)) \nonumber \\ 
     &+& i(m+n)\gamma_1(t)/2\Big] \mathrm{d}t
     \label{eq:phinm_soln}.
\end{eqnarray}

By tracing out the cavity mode in Eq.~\ref{eq:ansatz_rho} one can hence write the corresponding quantum channel of the geometric phase gate on a spin basis state as in Eq.~\eqref{eq:eff_channel}. 

\section{Optimal state-preparation-protocol at \texorpdfstring{$gT \rightarrow \infty$}{infinite gT} and at finite \texorpdfstring{$gT$}{gT}}
In this section, we discuss the numerical optimization details for both the cases of the cavity pulse duration in the application of $\mathcal{E}_{\mathrm{gpg}}$ corresponding to $gT \rightarrow \infty$ and for a finite $gT$. 

For finding the optimal state preparation protocol parameters for the case of $gT \rightarrow \infty$, we make use of Eq.~\eqref{eq:phase_T_infty}(in the main text) in the application of $\mathcal{E}_{\mathrm{gpg}}$ 
where we have $\phi = \frac{1}{\delta} \int_{0}^{T} |\zeta(t)|^2$, and hence we must have the same sign for $\phi_j$ and $\delta_j$ in each step $j$ while finding the optimal parameters. We hence perform a boundless optimization using $\varphi_{nm} = (n^2- m^2)\phi_j + (m-n)^2 \frac{i\kappa}{2} \left|\frac{\phi_j}{\delta_j}\right| + (m+n) \frac{i\gamma}{2g^2}\left|\phi_j \delta_j \right|$, and post adjust the sign of $\delta_j$ corresponding to the sign of $\phi_j$. 

For finding the optimal protocol parameters for a finite cavity pulse duration, the quantum channel $\mathcal{E}_{\mathrm{gpg}}$ from Eq.~\ref{eq:eff_channel} is applied using the solution in Eqs.\ref{eq:beta_soln}-~\ref{eq:phinm_soln} with $\beta_{n}(0)= 0$, that is assuming the cavity mode starts in vacuum(note that the protocol is independent of the initial cavity state, see Ref.~\cite{PhysRevA.110.062610}). The optimization is partially bounded where the bounds are introduced for the $\delta_{j}$ values arising from the physical constraint of limiting the pulse duration to $gT$ while keeping reasonable $\mathrm{max}_{t}|\eta(t)|$. The constraint can be explicitly written from the transformation from the full Hamiltonian to the effective Hamiltonian in Eq.~\eqref{eq:H_eff}, as $|\zeta|^2 < g^2/4$, which sets the bounds $\delta^{(j)} \in \left(\frac{2\pi}{T} , \frac{3g^2T}{32\phi^{(j)}}\right)$, see also Sec.~\ref{subsec:zeta_constraint_delta_bounds}. We start the optimization with the parameters corresponding to the $T \rightarrow \infty$ case, with $\delta^{(j)}$s adjusted within the bounds mentioned above. Fig.\ref{fig:N10_T_results} shows the obtained optimal $(\Delta \beta)^2_{\mathrm{GHZ}}$ and $(\Delta \beta)^2_{N/2}$ for $N=10$ as a function of the cavity pulse duration in each application of $\mathcal{E}_{\mathrm{gpg}}$. The obtained optimal values optimal $(\Delta \beta)^2_{\mathrm{GHZ}}$ and $(\Delta \beta)^2_{N/2}$ show a dependence on $\gamma/\kappa$ and is minimal for $\gamma= \kappa$. For large cooperativities of $C> 100$, the optimal values converge close to the values corresponding to  $gT\rightarrow \infty $ case for pulse durations $gT \approx 30-40 g^{-1}$.

\begin{figure}[hbt]
    \centering
    \includegraphics[width= 0.5\linewidth]{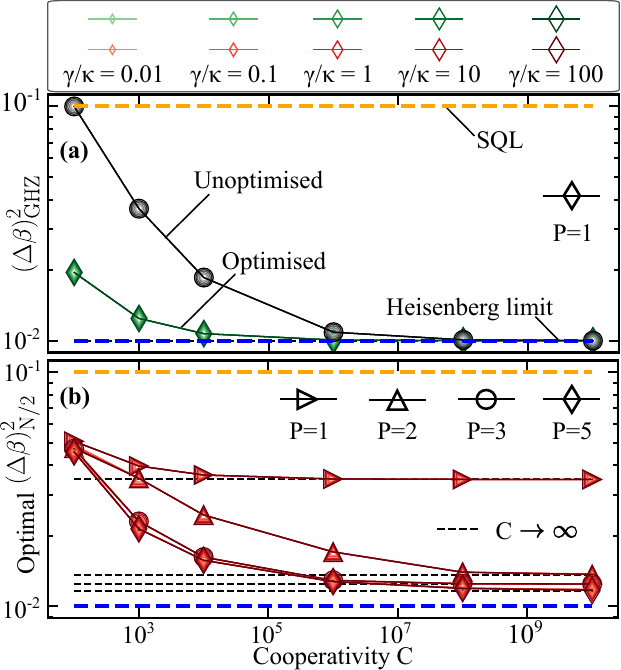}
    \caption{(a) Optimal $(\Delta \beta)^2_{\mathrm{GHZ}}$ for $P=1$ step obtained as a function of Cooperativity $C$ , plotted for $N= 10$ and different ratios $\gamma/\kappa$. The circle markers correspond to the results obtained with the application of \textit{unoptimised} pulses referring to the pulses which prepare the ideal GHZ state with $(\Delta \beta)^2_{\mathrm{GHZ}}= 1/N^2$ for the case $\kappa= \gamma=0$. (b) $(\Delta \beta)^2_{N/2}$ for $P=1, 2, 3, 5$ steps obtained as a function of Cooperativity $C$ , plotted for $N= 10$ and different ratios $\gamma/\kappa$. }
    \label{fig:N10_C_results}
\end{figure}

\begin{figure}[hbt]
    \centering
    \includegraphics[width= 0.5\linewidth]{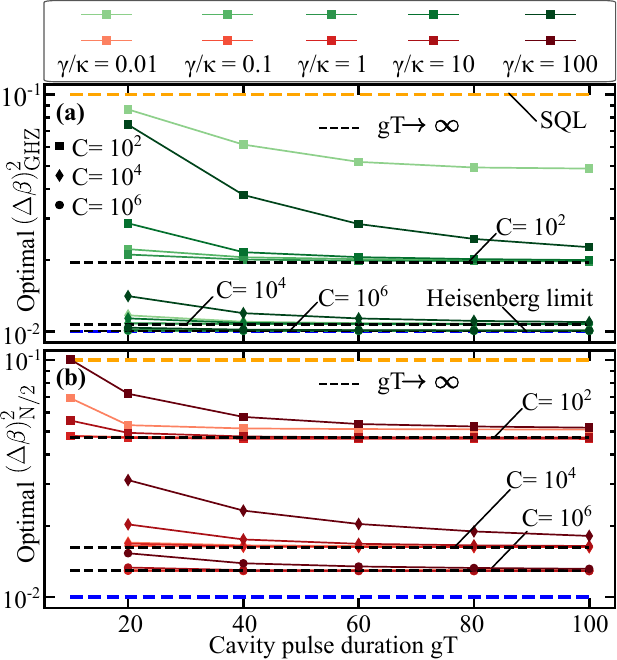}
    \caption{(a) Optimal $(\Delta \beta)^2_{GHZ}$ for $P=1$ step and (b) $(\Delta \beta)^2_{N/2}$ for $P=3$ steps obtained as a function of the cavity drive pulse duration $gT$, plotted for $N= 10$, cooperativities $C= 10^2, 10^4, 10^6$ and different ratios $\gamma/\kappa$.}
    \label{fig:N10_T_results}
\end{figure}

\section{Exact derivatives of the protocol cost function}

In this section we provide the derivatives of our protocol cost function $(\Delta \beta)^2$ with respect to all parameters $\Theta = \{ \theta^{\alpha}_0, \theta^{\beta}_0, \theta^{\gamma}_0, \theta^{\alpha}_j, \theta^{\beta}_j, \theta^{\gamma}_j, \phi_j, \delta_j \dots \forall j= 1, 2\dots P \} $. We first start out by writing the derivatives of the states obtained after each protocol step. 

Our protocol starts with the application of  $\hat{ \mathcal{U}}_0$ on the initial state $\rho_{\mathrm{in}}= \ket{\mathcal{D}_{0}}$, giving $\rho_{0} = \hat{U}_{0} \rho_{\mathrm{in}} \hat{U}_{0}^{\dagger}$. The states $\rho_{j}$ obtained after application of protocol step $j$ for $j= 1, 2 \dots P$ are obtained as
\begin{eqnarray}
    \rho_{j} &=& \hat{U}_{j} \mathcal{E}_{gpg}(\rho_{j-1}) \hat{U}_{j}^{\dagger}.
\end{eqnarray}

It is then straightforward to write the derivatives of $\rho_0$ and $\rho_{j}$ $\forall j= 1, 2, \dots P$ with respect to the parameters $\Theta$, which are obtained as given below
\begin{eqnarray}
    &\partial_{\theta_{0}}\rho_{0} &= (\partial_{\theta_{0}} \hat U_{0}) \rho_{\mathrm{in}} \hat U_{0}^{\dagger} + \hat U_{0} \rho_{\mathrm{in}} (\partial_{\theta_{0}^{\alpha}}\hat U_{0}^{\dagger}),\nonumber\\
    &\partial_{\theta_{j}}\rho_{0} &= \partial_{\phi_{j}}\rho_{0} = \partial_{\delta_{j}}\rho_{0} =0,\nonumber\\
\nonumber
    &\partial_{\theta_{j}}\rho_{j} &= (\partial_{\theta_{j}} \hat U_{j}) \mathcal{E}_{gpg}(\rho_{j-1}) \hat U_{j}^{\dagger} + \hat U_{j} \mathcal{E}_{gpg}(\rho_{j-1}) (\partial_{\theta_{j}}\hat U_{j}^{\dagger} ),\\
    \nonumber
    &\partial_{\phi_{j}} \rho_{j}&= \hat U_j (\sum_{n, m}  \partial_{\phi_j}(\varphi_{nm}) e^{i\varphi_{nm}} \bra{\mathcal{D}_n} \rho_{j} \ket{\mathcal{D}_m}
    \ket{\mathcal{D}_n} \bra{\mathcal{D}_m}) \hat U_j^{\dagger},\nonumber\\
    &\partial_{\delta_{j}} \rho_{j}&= \hat U_j ( \sum_{n, m}  \partial_{\delta_j}(\varphi_{nm}) e^{i\varphi_{nm}} \bra{\mathcal{D}_n} \rho_{j} \ket{\mathcal{D}_m}
    \ket{\mathcal{D}_n} \bra{\mathcal{D}_m} ) \hat U_j^{\dagger} \nonumber\\
    &\partial_{\Theta_{k<j}}\rho_{j} &=  \mathcal{E}_j \dots \mathcal{E}_{k+1} (\partial_{\Theta_{k}} \rho_{k}), \partial_{\Theta_{k>j}}\rho_{j} = 0 \nonumber.
\end{eqnarray}
We have used the shorthand $\theta_{j}$ for $\theta_{j}^{\alpha}, \theta_{j}^{\beta}, \theta_{j}^{\gamma}$ and $\Theta_{j}$ refers to all elements in the set $\{ \theta_{j}^{\alpha}, \theta_{j}^{\beta}, \theta_{j}^{\gamma}, \phi_{j}, \delta_{j} \}$ in the equations above. Note that the derivatives of the state $\rho_{j}$ are simply obtained by performing similar operations- applying the geometric-phase-gate operation $\mathcal{E}_{\mathrm{gpg}}$ and global spin rotation operations. For example, obtaining $\partial_{\phi_{j}} \rho_{j}$ and $\partial_{\delta_{j}} \rho_{j}$ are similar to calculating $\mathcal{E}_{\mathrm{gpg}}(\rho_{j})$ but with modified phases $e^{i\varphi_{nm}} \rightarrow \partial_{\phi_j}(\varphi_{nm}) e^{i\varphi_{nm}}$ and $e^{i\varphi_{nm}} \rightarrow \partial_{\delta_j}(\varphi_{nm}) e^{i\varphi_{nm}}$ respectively. 
The optimal probe state is prepared after at $j=P$ protocol steps which we denote by $\rho_{P}=\rho_{\opt}$. With the prescription described above we obtain the exact derivatives corresponding to $\partial_{\Theta}\rho_{\opt}$, for all parameters $\Theta$.
In the following, we obtain the derivatives of the protocol cost function $(\Delta \beta)^2$ for the two choices of the measurement operator $\hat M$ corresponding to $(\Delta \beta)^2_{\mathrm{GHZ}}$ and $(\Delta \beta)^2_{N/2}$ for case I and II below respectively.
\subsubsection{Case I: Choosing \texorpdfstring{$\hat M= \hat{\mathcal{P}}_{x}= \bigotimes_{i=1}^{N}\hat \sigma_{x}^{(i)}$}{M as parity operator along x}}
The operator $\mathcal{P}_{x}= \bigotimes_{i=1}^{N}\hat \sigma_{x}^{(i)}$ measures the parity of the state along $x$. Using $e^{i(\pi/2) \hat \sigma_{x}}= i \hat \sigma_{x}$, we rewrite $\hat M$ as 
\begin{eqnarray}
    \hat M &=& e^{i \pi (\hat J_x-  N/2)}.
\end{eqnarray}
We choose the field generator corresponding to a field along $z$ for this case as $\hat H_{\Vec{z}} = J \hat J_z$. Let the state obtained after the rotation of the optimal probe state $\rho_{\opt}$ by an angle $\beta= Jt$ along the field axis be denoted by $\rho_{\opt}^{\beta}$. We obtain 
\begin{eqnarray}
    &&\rho_{\opt}^{\beta} = e^{-i\beta \hat J_z }\rho_{\opt}e^{i\beta \hat J_z },  \hspace{0.1em} \partial_{\beta}\rho_{\opt}^{\beta} = i \left[\rho_{\opt}^{\beta}, \hat J_z \right],\\
    &&\partial_{\Theta}\partial_{\beta}\rho_{\opt}^{\beta}= i \left[\partial_{\Theta}\rho_{\opt}^{\beta}, \hat J_z \right],\\
    &&\langle \hat{\mathcal{P}}_{x} (\beta) \rangle = \mathrm{Tr}(\hat{\mathcal{P}}_{x}, \rho_{\opt}^{\beta}),\\
    &&\partial_{\Theta}\langle \hat{\mathcal{P}}_{x} (\beta) \rangle =\mathrm{Tr}(\hat{\mathcal{P}}_{x}, \partial_{\Theta}\rho_{\opt}^{\beta}),
\end{eqnarray}
where $\partial_{\Theta}\rho_{\opt}^{\beta} = e^{-i\beta \hat J_z }(\partial_{\Theta} \rho_{\opt})e^{i\beta \hat J_z }$. Similarly, $\langle \hat{\mathcal{P}}_{x}^2 (\beta) \rangle= \mathrm{Tr}(\hat{\mathcal{P}}_{x}^2, \rho_{\opt}^{\beta})$, $\partial_{\Theta}\langle \hat{\mathcal{P}}^2_{x} (\beta) \rangle= \mathrm{Tr}(\hat{\mathcal{P}}^2_{x}, \partial_{\Theta}\rho_{\opt}^{\beta})$,  $\partial_{\beta} \langle \hat{\mathcal{P}}_{x} (\beta) \rangle = \mathrm{Tr}(\hat{\mathcal{P}}_{x}, \partial_{\beta} \rho_{\opt}^{\beta})$ and $\partial_{\Theta}\partial_{\beta} \langle \hat{\mathcal{P}}_{x} (\beta) \rangle = \mathrm{Tr}(\hat{\mathcal{P}}_{x}, \partial_{\Theta}\partial_{\beta} \rho_{\opt}^{\beta})$. 

With these, we obtain the derivatives of $(\Delta \beta)^2_{\mathrm{GHZ}}$ as
\begin{eqnarray}\label{eq:del_J}
    &&\partial_{\Theta}(\Delta \beta)^2_{\mathrm{GHZ}} =\left[\left(\partial_{\Theta}\langle \hat{\mathcal{P}}^2_{x} (\beta) \rangle- \partial_{\Theta} (\langle \hat{\mathcal{P}}_{x} (\beta) \rangle )^2\right)\left|\partial_{\beta}  \langle \hat{\mathcal{P}}_{x} (\beta) \rangle \right|^2  \right. \nonumber \\
    &&\left. - \left(\langle \hat{\mathcal{P}}^2_{x} (\beta) \rangle- \langle \hat{\mathcal{P}}_{x} (\beta) \rangle ^2\right)  \partial_{\Theta}\left|\partial_{\beta}  \langle \hat{\mathcal{P}}_{x} (\beta) \rangle \right|^2\right]/ \left|\partial_{\beta}  \langle \hat{\mathcal{P}}_{x} (\beta) \rangle \right|^4.
\end{eqnarray}

\subsubsection{Case II: Choosing \texorpdfstring{$\hat M= \hat J_z^2$}{M as collective spin operator along z}}

For this choice of measurement operator $\hat M$, we choose the field along $y$ axis corresponding to $\hat H_{\Vec{y}} = J \hat J_y$. The second and the fourth moments for $\hat M= \hat J_z^2$ after rotation of the probe state $\rho_{\opt}$ by angle $\beta= Jt$, written with $\langle \hat X \rangle = \mathrm{Tr}(\hat X, \rho_{\opt})$ are given by 
\begin{align}
    \expect{\hat J_z^2(\beta)} &= \expect{\hat J_z^2} \cos^2{\beta} + \expect{\hat J_x^2}\sin^2{\beta} - \expect{\anticomm{\hat J_z}{\hat J_x}}\sin{\beta}\cos{\beta},\\
    \expect{\hat J_z^4(\beta)} &= \expect{\hat J_z^4} \cos^4{\beta} + \expect{\hat J_x^4}\sin^4{\beta}\\
    \nonumber
    &+ (\expect{\anticomm{\hat J_z} {\hat J_x}^2}+ \expect{\anticomm{\hat J_z^2}{\hat J_x^2}})\cos^2{\beta}\sin^2{\beta}\\
    \nonumber
    &- \expect{\hat A} \cos^3{\beta}\sin{\beta} - \expect{\hat B} \cos{\beta}\sin^3{\beta},
\end{align}
where $\hat A= \anticomm{\hat J_z^2}{\anticomm{\hat J_z}{\hat J_x}}$ and $\hat B= \anticomm{\hat J_x^2}{\anticomm{\hat J_z}{\hat J_x}}$.

The variance of the measurement results is then obtained as $(\Delta \hat J_z^2(\beta))^2= \expect{\hat J_z^4(\beta)} -  \expect{\hat J_z^2(\beta)} ^2$. The derivative term in the denominator of $(\Delta \beta)^2$ is obtained as 
\begin{align}
    \partial_{\beta} \expect{\hat J_z^2(\beta)} &= 2 (\expect{\hat J_x^2} - \expect{\hat J_z^2}) \cos{\beta}\sin{\beta}\\
    \nonumber
    &- \expect{\anticomm{\hat J_z}{\hat J_x}} (\cos^2{\beta}- \sin^2{\beta}).
\end{align}

By writing $\partial_{\Theta} \langle \hat X \rangle= \mathrm{Tr}(\hat X, \partial_{\Theta}\rho_{\opt})$, it is straightforward to obtain $\partial_{\Theta}(\Delta \beta)^2_{N/2}$ similar to Eq.~\eqref{eq:del_J}. 

\section{Optimal state preparation protocol parameters}

In this section, we tabulate the obtained optimal parameters $\Theta$ in $\mathcal{E}_{q}$ which prepare the optimal probe states $\rho_{\mathrm{opt}}$ minimising $(\Delta \beta)^2_{\mathrm{GHZ}}$ and $(\Delta \beta)^2_{N/2}$ in Tables \ref{tab:GHZ_values_table} and \ref{tab:Nb2_values_table} respectively. 

\begin{table}
    \centering
    \begin{tabular}{|c|c|c|c|c|}
    \hline
        $N$ & $C$ & $\gamma/\kappa$ & $N(\Delta \beta)^2_{\mathrm{GHZ}}$& $(\theta_{0}^{\alpha}, \theta_{0}^{\beta}, \theta_{0}^{\gamma})$\\
        &&&&$(\phi_1, \delta_1,\theta_{1}^{\alpha}, \theta_{1}^{\beta}, \theta_{1}^{\gamma})$, $\Delta_1$\\
        \hline
        \hline
        10 & $10^2$ & 0.01 & 0.61& (0.98, 1.57, 0.88)\\
        &&&&(0.86, 2.18g, -1.16, 1.57, 0.96), 26g\\
        \hline
        && 1.0 & 0.20& (1.57, 1.41, 0.34)\\
        &&&&(1.56, 0.48g, 0, 1.57, 1.36), 237g\\
        \hline
         & $10^4$ & 0.01 & 0.109& (0, 1.56, 0.50)\\
        &&&&(1.56, 2.15g, 0, 1.57, -0.04), 9g\\
        \hline
        && 1.0 & 0.107& (-0.22, 1.55, 0.36)\\
        &&&&(1.57, 0.44g,0, 1.57, 0.19), 267g\\
        \hline
        \hline
        40 & $10^2$ & 1.0 & 0.096& (-0.34, 1.14, 0.50)\\
        &&&&(1.61, 0.30g, 0, 1.57, 0.07), 457g\\
        \hline
         & $10^4$ & 0.01 & 0.030& (1.51, 1.54, 0.37)\\
        &&&&(1.57, 2.03g , 0.08, 1.57, 1.58), 12g\\
        \hline
        && 1.0 & 0.029& (-0.04, 1.53, 0.37)\\
        &&&&(1.57, 0.28g, 0.08, 1.57,0.02), 497g\\
        \hline
        \hline
        100 & $10^2$ & 0.01 & 0.15& (1.51, 0.98, 0.69)\\
        &&&&(1.56, 2.13g, 0, 1.57, 1.32), 10g\\
        \hline
        && 1.0 & 0.07& (1.47, 0.87, 0.69)\\
        &&&&(1.64, 0.24g, 0, 1.57, 1.29), 597g\\
        \hline
         & $10^4$ & 0.01 & 0.013& (1.42, 1.51, 0.22)\\
        &&&&(1.57, 1.80g, 0.03, 1.57, 1.66 ), 19g\\
        \hline
        && 1.0 & 0.013& (1.38, 1.51, 0.22)\\
        &&&&(1.57, 0.18g, 0.03, 1.57, 1.63), 844g\\
        \hline
    \end{tabular}
    \caption{Optimal state preparation protocol parameters $\Theta_{\opt}$ minimizing $(\Delta \beta)^2_{\mathrm{GHZ}}$. The listed values correspond to the cavity pulses in the application of geometric phase gate $\mathcal{E}_{\mathrm{gpg}}$ of duration of $T= 40g^{-1}$. The $\Delta_{j}$ values are derived from the optimal $\phi_j$, $\delta_j$ by inverting the pulse $\zeta(t)$ in Eq.~\eqref{eq:H_eff} to $\eta(t)$ in the full Hamiltonian (see \cite{inverting_zeta_note}, and Ref.\cite{PhysRevA.110.062610}). An extra rotation along $\hat z$ direction to set $\beta_{\opt}=0$ is incorporated in $\theta^{\alpha}_{1}$\cite{extra_rotation_note}.}
    \label{tab:GHZ_values_table}
\end{table}

\begin{table}
    \centering
    \begin{tabular}{|c|c|c|c|c|l|}
    \hline
        $N$ & $C$ & $\gamma/\kappa$ & $N(\Delta \beta)^2_{N/2}$& $(\theta_{0}^{\alpha}, \theta_{0}^{\beta}, \theta_{0}^{\gamma})$, $\theta_{-1}^{\beta}$\\
        &&&&$(\phi_1, \delta_1,\theta_{1}^{\alpha}, \theta_{1}^{\beta}, \theta_{1}^{\gamma})$, $\Delta_1$ \\
        &&&&$(\phi_2, \delta_2,\theta_{2}^{\alpha}, \theta_{2}^{\beta}, \theta_{2}^{\gamma})$, $\Delta_2$\\
        &&&&$(\phi_3, \delta_3,\theta_{3}^{\alpha}, \theta_{3}^{\beta}, \theta_{3}^{\gamma})$, $\Delta_3$ \\
        \hline
        \hline
        10 & $10^2$ & 0.01 & 0.51& (1.19, 2.10, 0.66), 2.22\\
        &&&&(0, -, -2.11, 1.50, 0.70), -\\
        &&&&(0.21, 8g, -2.46, 2.15, -1.72), 8g\\
        &&&&(0.02, 7.06g, -0.96, 2.86, -2.01), 39g\\
        \hline
        && 1.0 & 0.47& (1.15, -0.97, 0.43), 0.24\\
        &&&&(0.08, 1.17g, 0.49, 1.08, 0.92), 323g\\
        &&&&(-0.14, -0.83g, -0.25, 0.96, 0.92),  393g\\
        &&&&(0.06, 1.06g, -0.99, -0.59, 0.13), 427g\\
        \hline
         & $10^4$ & 0.01 & 0.165& (-0.08, -1.57, 0.63), 0.09\\
        &&&&(0.10, 7.22g, 0.29, 2.52, -0.48), 17g\\
        &&&&(0.30, 5.66g, 1.91, 0.18, -0.42), 11g\\
        &&&&(1.38, 1.86g, 0.92, 0, 1.57), 21g\\
        \hline
        && 1.0 & 0.163& (-0.34, 1.57, 0.42), 2.09\\
        &&&&(0.10, 1.14g, 1.68, 0.61, -0.23), 293g\\
        &&&&(0.29, 0.67g, 0.49, 0.18, 1.25), 362g\\
        &&&&(1.38, 0.25g, 3.14, 2.00, 0.79), 611g\\
        \hline
        \hline
        40 & $10^2$ & 0.01 & 0.21& (0.04, 0, 0.41), -0.03\\
        &&&&(0.95, 0.41g, -1.03, 1.51, 0.43), 265g\\
        &&&&(-0.07, -7.07g, -2.59, 0.63, -0.19), 21g\\
        &&&&(0.03, 7.07g, 1.82, -0.26,-2.74 ), 33g\\
        \hline
        && 1.0 & 0.20& (0.36, -1.51, 0.49), -1.08\\
        &&&&(0.06, 0.82g, -0.90, 2.47, 0.45), 600g\\
        &&&&(-0.04, -0.71g, -0.41, 0.56, -0.84), 900g\\
        &&&&(0.01, 0.94g, 0.33, 1.16, -0.90), 1200g\\
        \hline
         & $10^4$ & 0.01 & 0.081& (0.29, 1.57, 0.74), 0.51\\
        &&&&(0.05, 6.34g, 1.50, 0.42, -0.02), 32g\\
        &&&&(0.22, 5.72g, -0.35, 0.09, 0.97), 15g\\
        &&&&(1.47, 0.89g, 0, 0.48, -0.60), 90g\\
        \hline
        && 1.0 & 0.086& (0.26, 1.57, 0.74), 0.51\\
        &&&&(0.04, 5.30g, 1.28, 0.43, -0.05), 44g\\
        &&&&(0.21, 0.67g, -0.67, 0.09, 0.75), 400g\\
        &&&&(1.45, 0.17g, 0, -0.48, -0.92), 900g\\
        \hline
        \hline
        100 & $10^2$ & 0.01 & 0.133& (0.23, 0, 0.72), 2.16\\
        &&&&(0.32, 0.64g, -1.93, -1.58, 0.92), 300g\\
        &&&&(-0.04, -7.07g, 1.83, -0.34, -1.05), 28g\\
        &&&&(0.04, 3.26g, -0.11, 0.97, 1.33), 90g\\
        \hline
         & $10^4$ & 0.01 & 0.045& (-1.17, 1.55, 0.53), 0\\
        &&&&(0.02, 15.69g, 0.078, -0.34, -1.07), 11g\\
        &&&&(0.17, 4.55g, 0.84, 0.10, 0.80), 26g\\
        &&&&(-0.08, -7.07g, 1.07, 0.04, 0.88), 19g\\
        \hline
        && 1.0 & 0.048& (0.49, 1.59, 0.71), -0.07\\
        &&&&(0.03, 0.76g, -0.08, 0.32, 0.85), 1000g\\
        &&&&(0.18, 0.25g, 0.26, 0.05, 0.33), 1800g\\
        &&&&(0.43, 0.17g, 0.21, 0.08, 0.75), 1700g\\
        \hline
    \end{tabular}
    \caption{Optimal state preparation protocol parameters $\Theta_{\opt}$ minimizing $(\Delta \beta)^2_{N/2}$. The listed values correspond to the cavity pulses in the application of geometric phase gate $\mathcal{E}_{\mathrm{gpg}}$ of duration of $T= 40g^{-1}$.The $\Delta_{j}$ values are derived from the optimal $\phi_j$, $\delta_j$ by inverting the pulse $\zeta(t)$ in Eq.~\eqref{eq:H_eff} to $\eta(t)$ in the full Hamiltonian (see \cite{inverting_zeta_note}, and Ref.\cite{PhysRevA.110.062610}). The angles $\theta_{-1}^{\beta}$ refer to the extra rotation along the field axis $\hat y$ at the end of the protocol steps to set $\beta_{\opt}=0$ \cite{extra_rotation_note}.}
    \label{tab:Nb2_values_table}
\end{table}

\section{Spins under local homogeneous dephasing during state preparation \label{SEC::DEPHASING_STATE_PREP}}

In this section, we study the robustness of our state preparation protocol against the local homogeneous dephasing process. We consider the dephasing effects introduced as local homogeneous dephasing processes, which can be described as a \textit{collective} process\cite{PhysRevA.78.052101}, and we work in the collective Hilbert space $\mathcal{H}_{C}$ of dimension $\sum_{J= J_{\mathrm{min}}}^{J_{\mathrm{max}}} (2J+1)$ where $J_{\mathrm{max}}= N/2$ and $J_{\mathrm{min}}= (N$ mod  $2)/2 $. We study primarily the effects of the local homogeneous dephasing process during the application of the geometric phase gate $\mathcal{E}_{\mathrm{gpg}}$ and consider negligible dephasing during the fast global spin rotation operations. We perform the numerical calculations in the collective Hilbert space using the \textit{piqs} solver\cite{shammah2018open}. 

In our geometric phase gate protocol implemented during the state preparation protocol, we make use of the cavity mode coupled with the $\ket{1}\leftrightarrow \ket{e}$ transition with strength $g$, while the state $\ket{0}$ remains uncoupled. 
To add finite local homogeneous dephasing in the three-level system, we model the three level dephasing with the jump operators $\mathcal{A}_{\gamma_{\phi}^{e}}^{(j)}= \ket{e_j}\bra{e_j}$ and $\mathcal{A}_{\gamma_{\phi}^{1}}^{(j)}= \ket{1_j}\bra{1_j}$ corresponding to dephasing of states $\ket{e}$ and $\ket{1}$ with rates $\gamma_{\phi}^{e}$ and $\gamma_{\phi}^{1}$ respectively~\cite{li2012pure}. 
We include as before the cavity mode decay with rate $\kappa$ and the corresponding jump operator $\mathcal{A}_{\kappa} = \ah$.
The state $\rho$ in the original frame evolves according to $\dot \rho = -i \left[\hat H, \rho\right] + \mathbf{L}[\rho]$ 
where 

\begin{eqnarray}
    \mathbf{L}[\rho]&=& \kappa \mathbf{L}_{\kappa}[\rho] + \sum_{j=1}^{N} \left( \gamma_{\phi}^{1} \mathbf{L}_{\gamma_{\phi}^{1}}^{(j)}[\rho] + \gamma_{\phi}^{e} \mathbf{L}_{\gamma_{\phi}^{e}}^{(j)}[\rho]\right),
\end{eqnarray}
with $\mathbf{L}_{\alpha}[\rho]= \mathcal{A}_{\alpha} \rho \mathcal{A}_{\alpha}^{\dagger} - \frac{1}{2}\{ \mathcal{A}_{\alpha}^{\dagger} \mathcal{A}, \rho\}$.

We move from the original frame to the effective frame by performing two basis transformations. The first basis transformation acts only on the cavity subspace mapping $\rho \rightarrow \rho'$ and $\hat H \rightarrow \hat H'$ with $\mathbf{L}[\rho']= \mathbf{L}[\rho]$. The second basis transformation defined by $\hat U = \exp\left[\frac{\lambda}{2}\hat O \right]$ with $\hat O = -e^{i\mu}\hat S^++e^{-i\mu}\hat S^-$, where $\mu = \mathrm{arg}(\alpha)$ and $\lambda$ such that $\cos(\lambda)= \Delta/\sqrt{\Delta^2 + 4g^2|\alpha|^2}$\cite{PhysRevA.110.062610} acts on the qubit subspace alone which maps $\rho' \rightarrow \Tilde{\rho} = \hat U \rho ' \hat U^{\dagger}$. We hence obtain 

\begin{align}
\begin{split}
    \dot {\Tilde{\rho}}&=  \left(-i \hat U \hat H' \hat U^{\dagger} + \partial_t( (\lambda/2) \hat O )\right)\Tilde{\rho}\\ &+ \Tilde{\rho } \left(i \hat U \hat H' \hat U^{\dagger}+  \partial_t((\lambda/2) \hat O^{\dagger}) \right) + \hat U \mathbf{L}[\rho]\hat U ^{\dagger}\\
    &\equiv \dot \rho_{\eff} = -i \left[ \hat H_{\eff}, \Tilde{\rho} \right] + \mathcal{L}(\rho_{\eff}).
\end{split}
\end{align}

In the effective frame, we map $\Tilde{\rho} \rightarrow \rho_{\eff}$ where we restrict the dynamics only to the computational states $|0\rangle$ and $|1\rangle$, by assuming that we initially always start with a state with $n_e=0$, neglecting energy terms of the order $\mathcal{O}(\Delta)$, and coupling terms of the order $\mathcal{O}(g)$ between the states with energy difference diverging with $\Delta/g \rightarrow \infty$. We use $\hat U \hat H' \hat U^{\dagger} + i \partial_t( (\lambda/2) \hat O = \hat H_{\eff} + \mathcal{O}(g)(\hat S^+, \hat S^{-}) \approx \hat H_{\eff}$. We map similarly $\Tilde{\mathcal{L}}[\Tilde{\rho}]= \hat U \mathbf{L}[\rho]\hat U ^{\dagger} \rightarrow \mathcal{L}(\rho_{\eff})$. The transformed jump operators $\Tilde{\mathcal{A}^{(j)}} = \hat U \mathcal{A}^{(j)} \hat U^{\dagger}$ are obtained as

\begin{eqnarray}
    \Tilde{\mathcal{A}^{(j)}_{\gamma_{\phi}^{1}}} &=& \mathcal{A}^{(j)}_{\gamma_{\phi}^{1}} \frac{1}{2} \left( 1+ \sqrt{1- 4|\zeta|^2/g^2}\right) \nonumber \\ &&+  \mathcal{A}^{(j)}_{\gamma_{\phi}^{e}} \frac{1}{2}\left(1-\sqrt{1- 4|\zeta|^2/g^2} \right) \nonumber \\&&- (e^{i\mu} (\mathcal{A}^{(j)}_{\gamma})^{\dagger} + e^{-i\mu} (\mathcal{A}^{(j)}_{\gamma}))\frac{1}{2}(|\zeta|/g),\label{eq:jump_ops_transformed_1}\\
    \Tilde{\mathcal{A}^{(j)}_{\gamma_{\phi}^{e}}} &=& \mathcal{A}^{(j)}_{\gamma_{\phi}^{e}} \frac{1}{2} \left( 1+ \sqrt{1- 4|\zeta|^2/g^2}\right) \nonumber \\&&+ \mathcal{A}^{(j)}_{\gamma_{\phi}^{1}} \frac{1}{2}\left(1-\sqrt{1- 4|\zeta|^2/g^2} \right) \nonumber \\&&+ (e^{i\mu} (\mathcal{A}^{(j)}_{\gamma})^{\dagger} + e^{-i\mu} (\mathcal{A}^{(j)}_{\gamma}))\frac{1}{2}(|\zeta|/g)
    \label{eq:jump_ops_transformed_2}
\end{eqnarray}

The Lindbladian $\mathcal{L}[\rho_{\eff}] $ is obtained from $\hat U \mathbf{L} \hat U^{\dagger}$ after applying similar assumptions described above as in the derivation of $\hat H_{\eff}$, given by
\begin{align}
\begin{split}
    &\mathcal{L}[\rho_{\eff}]= \kappa \mathcal{L}_{\kappa}[\rho_{\eff}]+ \sum_{j=1}^{N}\left(\gamma'_{\phi}\mathcal{L}_{\gamma'_{\phi}}^{(j)}[\rho_{\eff}] + \gamma'\mathcal{L}_{\gamma'}^{(j)}[\rho_{\eff}]\right),\\
    &\mathcal{L}_{\kappa}[\rho_{\eff}] = \mathbf{L}_{\kappa}[\rho_{\eff}],\\
    &\mathcal{L}_{\gamma'_{\phi}}^{(j)}[\rho_{\eff}] =  \mathcal{A}_{\gamma'_{\phi}}^{(j)}\rho_{\eff} (\mathcal{A}_{\gamma'_{\phi}}^{(j)})^{\dagger} - \frac{1}{2}\{(\mathcal{A}_{\gamma'_{\phi}}^{(j)})^{\dagger} \mathcal{A}_{\gamma'_{\phi}}^{(j)} ,\rho_{\eff}\},\\
    &\mathcal{L}_{\gamma'}^{(j)}[\rho_{\eff}]= -\frac{1}{2} \{\hat n_{1}^{(j)},\rho\},
\end{split}
\label{eqs:L_eff_collective}
\end{align}

where 

\begin{eqnarray}
    \gamma'_{\phi} &=& \gamma_{\phi}^{1} \frac{(1+ \sqrt{(1-4|\zeta|^2/g^2)})^2}{4}+ \gamma_{\phi}^{e} \frac{(1- \sqrt{(1-4|\zeta|^2/g^2)})^2}{4}, \nonumber \\
    \mathcal{A}^{(j)}_{\gamma'_{\phi}}&=& \frac{1}{2}\sigma_{z}^{(j)},\hspace{0.5em} \gamma' = (\gamma_{\phi}^1 + \gamma_{\phi}^e) \frac{|\zeta|^2}{g^2}.
\end{eqnarray}

We combine  $\mathcal{L}_{\gamma'}^{(j)}[\rho_{\eff}]$ in the Hamiltonian as non-hermitian contribution resulting in solving the system with
\begin{eqnarray}
    \hat H_{\eff} &=& \delta \ah^\dagger\ah +  \left(-i\frac{(\gamma_1+ \gamma')}{2} + \zeta \ah^\dagger+ \zeta^* \ah\right)\hat{n}_1\nonumber,\\
    \mathcal{L}[\rho_{\eff}] &=& \kappa \mathcal{L}_{\kappa}[\rho_{\eff}]+ \gamma'_{\phi}\sum_{j=1}^{N}\mathcal{L}_{\gamma'_{\phi}}^{(j)}[\rho_{\eff}]
    \label{eq:eff_model_deph_in_state_prep}
\end{eqnarray}

\begin{figure}
    \centering
    \includegraphics[width=0.5\linewidth]{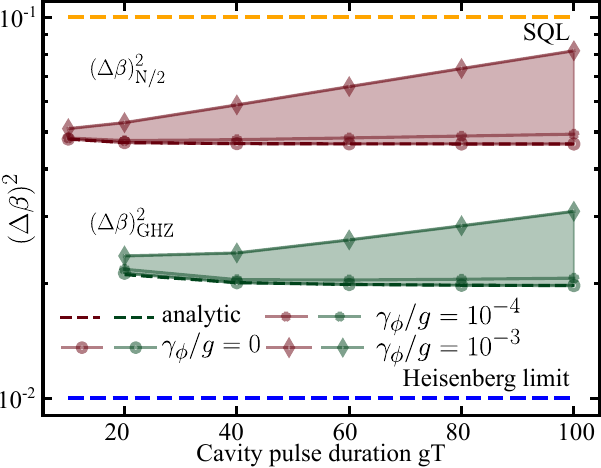}
    \caption{$(\Delta \beta)^2_{N/2}$ and $(\Delta \beta)^2_{\mathrm{GHZ}}$ obtained with our optimal state preparation protocol when local homogeneous dephasing of states $\ket{1}$ and $\ket{e}$ is added with rates $\gamma_{\phi}^{1}= \gamma_{\phi}^{e}= \gamma_{\phi}$, for $N=10$,  $C= 10^2$, $\gamma/\kappa=1.0$. The  markers correspond to numerical results obtained with simulations performed with the effective model derived in Eq.~\eqref{eq:eff_model_deph_in_state_prep} in the collective Hilbert space. The dashed lines correspond to the values obtained with analytical solutions in Eqs.~\eqref{eq:beta_soln}-\eqref{eq:phinm_soln}. }
    \label{fig:state_prep_dephasing}
\end{figure}

In Fig.~\ref{fig:state_prep_dephasing}, $(\Delta \beta)^2_{N/2}$ and $(\Delta \beta)^2_{\mathrm{GHZ}}$ is plotted by simulating the master equation dynamics with the model described above (solid lines) with dephasing rates $\gamma_{\phi}^{1}= \gamma_{\phi}^{e} = \gamma_{\phi}= 0, 10^{-4}\,g, 10^{-3}\,g$ for $N=10$, $C= 10^2$, $\gamma/\kappa= 1.0$. The results with $\gamma_{\phi}/g=0$(circle markers) coincide with the results obtained with analytical solution(dashed lines) in Eqs.~\eqref{eq:beta_soln}-\eqref{eq:phinm_soln}, which validate our state preparation protocol. We see that the optimal probe states remain quite robust against dephasing rates of the order $\gamma_{\phi}/g < 10^{-4}$.

\section{Spins under local homogeneous dephasing during signal collection}

In this section, we discuss the performance of the prepared optimal probe states during signal collection in a field-sensing experiment where spin qubits are additionally subjected to homogeneous local dephasing. Homogeneous local dephasing can be described as a \textit{collective} process \cite{PhysRevA.78.052101}, with each $N$-qubit state $\rho$ in the  collective Hilbert space $\mathcal{H}_{C}$ of dimension $\sum_{J= J_{\mathrm{min}}}^{J_{\mathrm{max}}} (2J+1)$, with $J_{\mathrm{max}}= N/2$ and $J_{\mathrm{min}}= (N$ mod  $2)/2 $. Having the field generator $\hat H_{\Vec{n}}= J \hat J_{\Vec{n}}$, where $J$ is the coupling strength of the spins with the field, we describe the homogeneous local dephasing (fluctuations in transition frequency) with rate $\gamma_{\phi}$ on the two-level spin with index '$j$' using the jump operator $\mathcal{A}^{(j)} = \frac{1}{2}\hat \sigma_{z}^{(j)}$ where $\hat \sigma_{z}^{(j)} = |1_j\rangle \langle1_j| - |0_j\rangle \langle 0_j|$. The optimal probe state $\rho_{\opt}$  evolves according to $\dot \rho_{\opt} = -i \left[\hat H_{\Vec{n}}, \rho_{\opt}\right] + \gamma_{\phi}  \sum_{j=1}^{N} \mathcal{L}^{(j)}[\rho_{\opt}]$, where $\mathcal{L}^{(j)}[\rho_{\opt}]= \mathcal{A}^{(j)} \rho_{\opt} (\mathcal{A}^{(j)})^{\dagger} - \frac{1}{2} \{(\mathcal{A}^{(j)})^{\dagger} \mathcal{A}^{(j)}, \rho_{\opt}\}$. We solve the model numerically using \textit{piqs} package~\cite{shammah2018open}, using the prepared optimal probe states as initial states at $Jt=0$. 

In the following, we first present the analytic expressions for the measured $(\Delta \beta)^2$ with ideal GHZ and $\ket{\Dicke{N}{N/2}}$ states in the presence of local homogeneous dephasing of the spins as a function of time during signal acquisition in field sensing experiment, and further present the numerical results of the measured $(\Delta \beta)^2$ with the optimal probe states prepared with different $N$ at finite cooperativities. All results are summarized in Fig.\ref{fig:signal_collection_sm}. 

\subsection{GHZ-like states undergoing dephasing during signal collection}

Analytical results of evolution of an ideal GHZ state acted upon by a field $\hat H_{\Vec{z}} = J \hat J_z$ in the presence of local homogeneous dephasing on the spins are presented in Ref.~\cite{shaji2007qubit}. We summarize the results here, and write the analytic expression for $(\Delta \beta)^2$ of ideal GHZ states (rotated by $\pi/2N$ along $z$ such that $\beta_{\opt}=0$).

In accordance with the definition of jump operator $\mathcal{A}_{j}= \frac{1}{2}\hat \sigma_{z}$ in the master equation dynamics(see main text), the local dephasing map on a single spin is defined as $\mathcal{A}_{t} (\hat \sigma_{x} \pm i \hat \sigma_{y}) = e^{-i \frac{\gamma_{\phi}}{2} t} e^{\mp i Jt } (\hat \sigma_{x} + i \hat \sigma_{y})$, and $\mathcal{A}_{t} (\hat \sigma_{z})= \hat \sigma_{z}$,  $\mathcal{A}_{t} (\hat {\mathbb{I}})= \hat {\mathbb{I}}$. 
This map can be directly applied on the ideal GHZ state expanded as~\cite{shaji2007qubit}
\begin{eqnarray}
\nonumber
    \rho_{GHZ} = \frac{1}{2^{N+1}} \left(\otimes_{j=1}^{N} (\hat{\mathbb{I}}+ \hat \sigma_{z;j}) + \otimes_{j=1}^{N} (\hat{\mathbb{I}}- \hat \sigma_{z;j})+ \otimes_{j=1}^{N} (\hat\sigma_{x;j} + i \hat \sigma_{y;j}) + \otimes_{j=1}^{N} (\hat\sigma_{x;j} - i \hat \sigma_{y;j})\right)
\end{eqnarray}
Now for the GHZ state rotated by $\pi/2N$ along $\hat z$ given by $e^{-i(\pi/2N)\hat J_z} \rho_{GHZ} e^{i(\pi/2N)\hat J_z} $, and under the dephasing map, for $\hat M= \mathcal{P}_{x}$ we obtain
\begin{align}
    \langle \hat M \rangle &= e^{-N\frac{\gamma_{\phi}}{2}t} \cos{(NJt+ \pi/2)},\quad ( \Delta \hat M )^2 = 1- e^{-N\gamma_{\phi} t}\cos^2{(NJt+ \pi/2)},\\
    \label{eq:variance_ideal_GHZ}(\Delta Jt)^2 &= \frac{e^{N\gamma_{\phi} t}}{N^2}\frac{1- e^{-N\gamma_{\phi} t}\cos^2{(NJt+ \pi/2)}}{|\sin{(NJt+ \pi/2)} + (\gamma_{\phi}/2J) \cos{(NJt+ \pi/2)}|^2}.
\end{align}

For the obtained noisy GHZ-like optimal probe states, we fit $\langle \hat M \rangle$ with
\begin{align}
    \label{eq:GHZ_like_exp_M_ansatz}
    \langle \hat M \rangle &= \sum_{m=N/2}^{1} \alpha_{m} e^{-m\gamma_{\phi}t}\cos{(2m(Jt+ \pi/(2N)))} + \alpha_{0},
\end{align}
and expect a similar $\sim \exp{N\gamma_{\phi}t}$ scaling in $(\Delta Jt)^2$. For the case of $\gamma_{\phi}/J= 0.01$ in Fig.\ref{fig:signal_collection}, we obtain non-zero fit parameters $\alpha_{N/2-1} = 0.78$, $\alpha_{N/2-2} = 0.16$, $\alpha_{0}= 0.05$. 

\subsection{\texorpdfstring{$\ket{\Dicke{N}{N/2}}$}{Dicke N/2}-like states undergoing dephasing during signal collection}
We can perform a similar calculation to evaluate the effect of dephasing during signal accumulation on the $\ket{\Dicke{N}{N/2}}$ state by a field $\hat H_{\Vec{y}} = J \hat J_y$. In this scenario, the map generated by the signal and that due to dephasing in the $\hat{z}$ basis do not commute. To simplify this calculation, we assume that the input state is a perfect $\ket{\Dicke{N}{N/2}}$, which then undergoes dephasing at a rate $\gamma_{\phi}$ over a time $t$, followed by perfect rotation of the system by the unitary $U=e^{-iJt \hat{J}_y}$ without dephasing. This models a field profile where the field strength $J(\tau)$ is near zero until time $t$ where it turns on strongly so that the integrated action angle is $\beta=\int_0^{t} J(\tau)d\tau=Jt$.
The variance of the estimation of $\beta$ given the measurement operator $\hat{M}=\hat{J}_z^2$ is given by \cite{toth_quantum_2014}:
\begin{equation}
\label{eq:varDicke}
(\Delta \beta)^2=\frac{(\Delta\hat{J}_x^2)^2f(\beta)+4\langle \hat{J}_x^2\rangle-3\langle \hat{J}_y^2\rangle-2\langle\hat{J}_z^2\rangle(1+\langle\hat{J}_x^2\rangle)+6\langle \hat{J}_z\hat{J}_x^2\hat{J}_z\rangle}{4(\langle \hat{J}_x^2\rangle-\langle \hat{J}_z^2\rangle)^2}
\end{equation}
where
\[
f(\beta)=\frac{(\Delta\hat{J}_z^2)^2}{\tan^2(\beta)(\Delta\hat{J}_x^2)^2}+\tan^2(\beta).
\]
Now we define the set of $n$ bit strings with Hamming weight $w$ as $\mathcal{B}^n_w=\{\vec{x}|\sum_j x_j=w\}$ and furthermore the distance between two binary strings as $d(\vec{x},\vec{y})=\sum_j |x_j-y_j|$. The Dicke state can be written
\[
\ket{\Dicke{N}{N/2}}=\sqrt{\frac{1}{\binom{N}{N/2}}}\sum_{\vec{x}\in \mathcal{B}^n_w}\ket{\vec{x}}\otimes \ket{\Dicke{N-n}{N/2-w}}\sqrt{\binom{N-n}{N/2-w}}.
\]
Let the output of dephasing map after time $t$ acting on a state $\rho$ be written $S_t(\rho)$.
Notice that the expression for the variance in Eq.~\eqref{eq:varDicke} involves second and fourth moments of angular momentum operators. This fact together with the permutation invariance property of the Dicke states, and the local action of the dephasing map, implies that the we can focus on the action of the map on a decomposition of the input state into a partition of the state into a subsystem of the first two or four qubits and the rest. Specifically we have the following decomposition of the output state: 
\[
\begin{array}{lll}
S_t\Big(\ket{\Dicke{N}{N/2}}\bra{\Dicke{N}{N/2}}\Big)&=&\frac{1}{\binom{N}{N/2}}\sum_w\sum_{\vec{x},\vec{y}\in\mathcal{B}^j_w}\ket{\vec{x}}\bra{\vec{y}}e^{-d(\vec{x},\vec{y})\gamma_{\phi}t}\otimes S_t\Big(\ket{\Dicke{N-j}{N/2-w}}\bra{\Dicke{N-j}{N/2-w}}\Big){\binom{N-j}{N/2-w}}\\
&&+\Big({\rm terms\ having\ } S_t\Big(\ket{\Dicke{N-j}{N/2-w}}\bra{\Dicke{N-j}{N/2-w'}}\Big){\rm with\ }w\neq w'\Big).
\end{array}
\]
where we can focus on this decomposition for $j=2,4$.
The last terms which are off diagonal in the Dicke basis will not contribute to expectation values of weight $2$ or $4$ Pauli operators, when we take the trace, namely $\langle \hat{O}\rangle=\Tr[\hat{O}S_t\Big(\ket{\Dicke{N}{N/2}}\bra{\Dicke{N}{N/2}}\Big)]$. 
The input state is invariant under rotations about $\hat{z}$ as is the dephasing map so
$\langle \hat{J}_x^2\rangle=\langle \hat{J}_y^2\rangle$.
Also because there are an equal number of diagonal terms with even and odd Hamming weight we have $\langle \hat{J}_z^2\rangle=\langle \hat{J}_z^4\rangle=0$. Now we write
\[
\hat{J}_x^2=\frac{1}{4}\sum_{j\neq k}X_jX_k+\frac{N}{4}{\bf 1}, \quad \hat{J}_x^4=\frac{1}{16}\sum_{j\neq k,j'\neq k'}X_jX_kX_{j'}X_{k'}+\frac{N}{8}\sum_{j\neq k}X_jX_k+\frac{N^2}{16}{\bf 1}.
\]
For $\langle \hat{J}_x^2\rangle$ the two point expectation value  
\[
\langle X_jX_k\rangle =\frac{e^{-2\gamma_{\phi}t} N}{2(N-1)}
\]
for $j\neq k$, of which there are $N(N-1)$ terms,
and hence 
\[
\langle \hat{J}_x^2\rangle=\frac{1}{4}\Big(\frac{e^{-2\gamma_{\phi}t}N^2}{2}+N\Big).
\]
For $\langle \hat{J}_x^4\rangle$, the four point expectation value 
\[
\langle X_jX_kX_{\ell}X_m\rangle =\frac{e^{-4\gamma_{\phi}t} 3N(N-2)}{8(N-1)(N-3)},
\]
for $j\neq k\neq \ell\neq m$, of which there are $N(N-1)(N-2)(N-3)$ terms.
The number of terms involving $\langle {\bf 1}\rangle$ are $3N^2-2N$. The remaining terms only involve two point expectation values $\langle X_jX_k\rangle$ with $j\neq k$ and there are $N^4-N(N-1)(N-2)(N-3)-(3N^2-2N)$ of them. Hence
\[
\langle \hat{J}_x^4\rangle=\frac{1}{16}\Big((3N^2-2N)+e^{-2 \gamma_{\phi}t}(3N^3-4N^2)+ \frac{e^{-4 \gamma_{\phi}t}3 N^2(N-2)^2}{8}\Big).
\]
Finally, we find
\[
\begin{array}{lll}
\langle \hat{J}_z\hat{J}_x^2\hat{J}_z\rangle&=&\Tr[\hat{J}_z \hat{J}_x^2\hat{J}_z S_t\Big(\ket{\Dicke{N}{N/2}}\bra{\Dicke{N}{N/2}}\Big)]\\
&=&\Tr[ \hat{J}_x^2\hat{J}_z S_t\Big(\ket{\Dicke{N}{N/2}}\bra{\Dicke{N}{N/2}}\Big)\hat{J}_z]\\
&=&\Tr[ \hat{J}_x^2 S_t\Big(\hat{J}_z\ket{\Dicke{N}{N/2}}\bra{\Dicke{N}{N/2}}\hat{J}_z\Big)]\\
&=&0
\end{array}
\]
using the fact that $\hat{J}_z$ commutes with the dephasing channel, and $\hat{J}_z\ket{\Dicke{N}{N/2}}=0$.
Hence we arrive at for $\beta=Jt$
\begin{equation}
\label{eq:variance_ideal_D_N_by_2}
(\Delta Jt)^2=\frac{16 e^{2\gamma_{\phi}t}(2e^{2\gamma_{\phi}t}+N)+(16 e^{4\gamma_{\phi}t}(N-1)+16 e^{2\gamma_{\phi}t}N(N-2)+N (12-12N+N^2))\tan^{2}(Jt) }{8N(2e^{2\gamma_{\phi}t}+N)^2}.
\end{equation}
Notice as expected, at $t=0$ the variance $(\Delta \beta)^2=\frac{2}{N(N+2)}$. 

\begin{figure}
    \centering
    \includegraphics[width=0.9\linewidth]{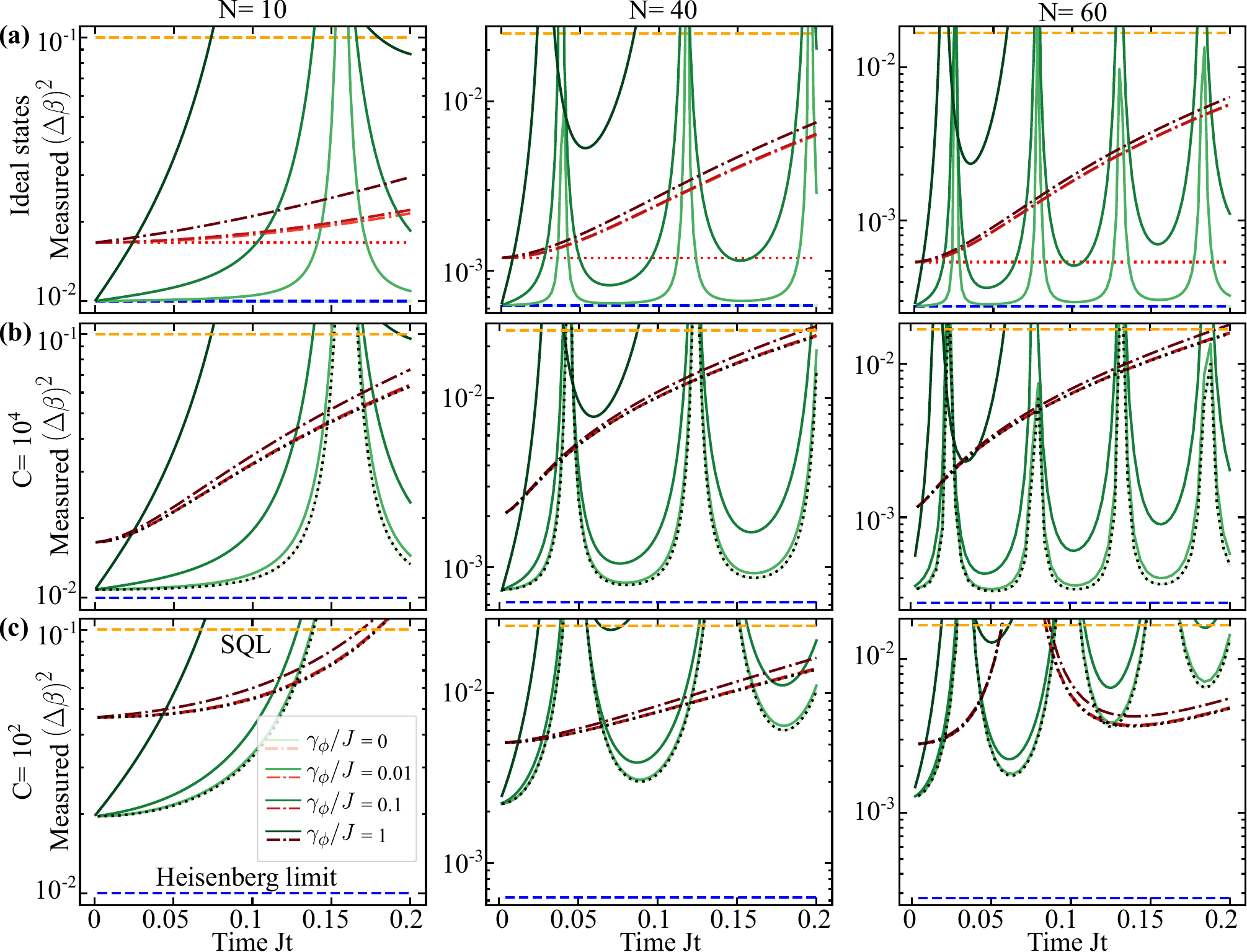}
    \caption{(a) $(\Delta \beta)^2= (\Delta Jt)^2$ as a function of dimensionless signal acquisition time $Jt$ for ideal GHZ states and ideal $\ket{\Dicke{N}{N/2}}$ states evolving under a field coupled with strength $J$ with local homogeneous dephasing acting on spins obtained in Eqs.~\eqref{eq:variance_ideal_GHZ} and ~\eqref{eq:variance_ideal_D_N_by_2} respectively, for dephasing rates $\gamma_{\phi}/J= 0,0.01,0.1,1.0$ and $N= 10, 40, 60$. The dotted red lines correspond to $(\Delta \beta)^2= 2/(N(N+2)$. (b) Measured $(\Delta \beta)^2$ as a function of dimensionless signal acquisition time $Jt$ by numerically evolving the optimal probe states prepared at cooperativity $C= 10^4$, $\gamma/\kappa= 1.0$ under a field coupled with spins with coupling strength $J$, with local dephasing on spins with rates $\gamma_{\phi}$ (for similar values as in panel (a)). Dotted black curves are the optimal $(\Delta \beta)^2$ obtained with analytic solution of $\mathcal{E}_{\mathrm{gpg}}$ for $\gamma_{\phi}/J=0$. (c) Similar to panel (b) for optimal states prepared at cooperativity $C= 10^2$, $\gamma/\kappa= 1.0$. The cooperativity $C$ values corresponding to an entire panel (row) and $N$ values corresponding to an entire column are indicated to the left and the top sides respectively. 
    Throughout, green solid lines (darker shades for larger $\gamma_{\phi}$) correspond to GHZ-like states while red dash-dot lines correspond to $\ket{\Dicke{N}{N/2}}$ states.}
    \label{fig:signal_collection_sm}
\end{figure}

In Fig.\ref{fig:signal_collection_sm}, we plot the measured $(\Delta \beta)^2= (\Delta Jt)^2$ as a function of the signal acquisition time $Jt$ for ideal GHZ and $\ket{\Dicke{N}{N/2}}$ probe states(panel (a)) with local homogeneous dephasing rates $\gamma_{\phi}/J = 0, 0.01, 0.1, 1.0$ for $N= 10, 40, 60$ and compare their performance in field-sensing experiment against the performance of the optimal probe states(similar to Fig.\ref{fig:signal_collection} but for longer signal collection times) prepared for $C= 10^4$, $\gamma/\kappa=1.0$(panel (b)) and $C= 10^2$, $\gamma/\kappa=1.0$(panel (c)). We observe a qualitatively similar behaviour of the optimal probe states prepared at finite cooperativities. 

\section{Local homogeneous spontaneous emission treated as a collective process}
\begin{figure}
    \centering
    \includegraphics[width=0.5\linewidth]{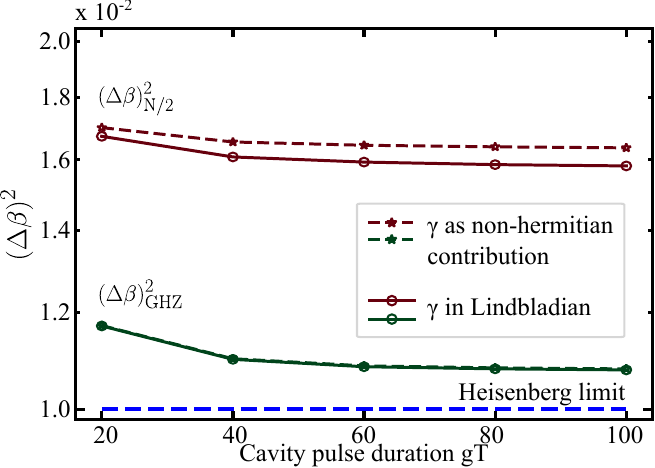}
    \caption{$(\Delta \beta)^2_{N/2}$ and $(\Delta \beta)^2_{\mathrm{GHZ}}$ obtained with our optimal state preparation protocol for $N=10$, $C= 10^4$, $\gamma/\kappa= 0.01$ with the spontaneous emission from the $\ket{e}$ state treated as a non-hermitian contribution (dashed lines, star markers) compared with the values obtained when decay is treated as a Lindbladian jump operator in the master equation formalism (solid lines, circle markers).}
    \label{fig:gamma_in_L_sm}
\end{figure}

In this section, we treat the local homogeneous spontaneous emission rate $\gamma$ of state $|e\rangle$ in the master equation approach with jump operator $\mathcal{A}_{\gamma}^{(j)}= \ket{1_j}\bra{e_j}$. 
 The transformed jump operator $\Tilde{\mathcal{A}_{\gamma}^{(j)}} = \hat U \mathcal{A}_{\gamma}^{(j)} \hat U^{\dagger}$ (similar to qubit basis transformation performed in Eqs.~\eqref{eq:jump_ops_transformed_1}-\eqref{eq:jump_ops_transformed_2}) is obtained as
\begin{eqnarray}
    \Tilde{\mathcal{A}^{(j)}_{\gamma}} &=&\mathcal{A}^{(j)}_{\gamma} \frac{1}{2} \left( 1+ \sqrt{1- 4|\zeta|^2/g^2}\right) \nonumber \\&& - (\mathcal{A}^{(j)}_{\gamma_{\phi}^{e}}- \mathcal{A}^{(j)}_{\gamma_{\phi}^{1}}) e^{i\psi}  (|\zeta|/g) \nonumber\\&&- (\mathcal{A}^{(j)}_{\gamma})^{\dagger}\frac{e^{i2\psi}}{2}\left(1- \sqrt{1- 4|\zeta|^2/g^2}\right).
\end{eqnarray}
We obtain a similar effective Lindbladian $\mathcal{L}_{\eff}$ in the same form as in Eqs.~\eqref{eqs:L_eff_collective}, with
\begin{eqnarray}
    \gamma_{\phi}'= \gamma |\zeta|^2/g^2, \hspace{1em} \mathcal{A}_{\gamma_{\phi}'}^{(j)}= \frac{1}{2}\sigma_{z}^{(j)},\\
    \gamma'= \gamma \frac{(1- \sqrt{(1- 4|\zeta|^2/g^2)})^2}{4}.
\end{eqnarray}
The effective model is reduced to
\begin{eqnarray}
    \hat H_{\eff} &=& \delta \ah^\dagger\ah +  \left(-i\frac{\gamma'}{2} + \zeta \ah^\dagger+ \zeta^* \ah\right)\hat{n}_1,\\
    \mathcal{L}[\rho_{\eff}] &=& \kappa \mathcal{L}_{\kappa}[\rho_{\eff}]+ \gamma'_{\phi}\sum_{j=1}^{N}\mathcal{L}_{\gamma'_{\phi}}^{(j)}[\rho_{\eff}].
\end{eqnarray}
In Fig.~\ref{fig:gamma_in_L_sm}, $(\Delta \beta)^2_{N/2}$ and $(\Delta \beta)^2_{\mathrm{GHZ}}$ for $N=10$, $C= 10^4$, $\gamma/\kappa=0.01$ is plotted by simulating the master equation dynamics with the model described above (solid lines). It is compared against the values obtained when $\gamma$ is treated as a non-hermitian contribution (dashed lines, model described in the main text, see Eq.~\eqref{eq:H_eff}). We see that the solid lines always lie very close or below the dashed lines, hence implying an upper bound on the variance corresponding to the $(\Delta \beta)^2$ values obtained in the main text by treating $\gamma$ as a non-hermitian contribution.

\section{Numerical validation with Master equation simulation with full Hamiltonian in full Hilbert space}

In this section, we validate our protocol by simulating the time-dependent open system dynamics by working in the full Hilbert space, with inhomogeneous coupling and by including $\gamma$ in the Master equation, that is by exactly simulating the following Lindblad master equation in the full Hilbert space. 
\begin{eqnarray}
    \dot \rho &=& - i \left[ \hat H , \rho \right] + \hat L \rho \hat L^{\dagger} - \frac{1}{2} \{ \hat L^{\dagger} \hat L , \rho \} \nonumber\\
    &+& \sum_{j=1}^{N} \left( \hat L^{(j)}_{\gamma} \rho (\hat L^{(j)}_{\gamma})^{\dagger} - \frac{1}{2} \left\{ (\hat L^{(j)}_{\gamma})^{\dagger} \hat L^{(j)}_{\gamma} , \rho \right \} \right) \label{eq:full_Master_equation_gamma},
\end{eqnarray}
where $\hat H = \Delta \hat n_e + \sum_{j=1}^{N} \left[(g_j |1_j\rangle \langle e_j|  + i\eta(t)) \ah + {\rm h.c} \right]$, $\hat L = \sqrt{\kappa} \ah$, $\hat L_{\gamma}^{(j)} = \sqrt{\gamma} |1_j\rangle \langle e_j|$. In our simulations, we choose a physically motivated inhomogeneous coupling strength profile given by $g_j = g/\sqrt{1 + (z_j/z_R)^2}$, where $z_j$ denotes the position of the $j$th spin qubit along the cavity axis chosen such that the spins are assumed to be placed at a spacing of $\approx 5\,\mu$m symmetrically along the cavity origin, and $z_R$ is the Rayleigh range of the cavity divergence which we set as $z_R = 130\mu$m. 

\begin{figure}
    \centering
    \includegraphics[width=0.5\linewidth]{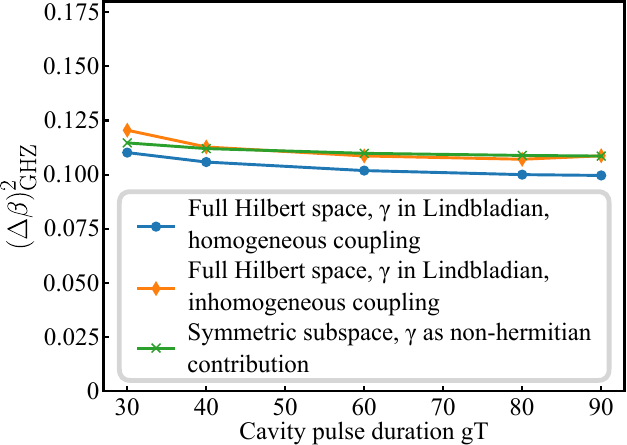}
    \caption{$(\Delta \beta)^2_{\rm GHZ}$ as a function of the cavity pulse duration $gT$ applied during the geometric phase gate protocol step for $N=4$, $C =100$, $\gamma/\kappa =0.1$. The cross markers (in green) correspond to the results obtained by our optimal state preparation protocol using the analytic solution of the geometric phase gate dynamics. The circle markers (in blue) and diamond markers (in orange) correspond to the results from simulations carried out in the full Hilbert space, incorporating $\gamma$ decay via the Lindbladian Master equation (see Eq.~\eqref{eq:full_Master_equation_gamma}), for the cases of homogeneous and inhomogeneous couplings (see text), respectively.}
    \label{fig:full_H_numerics_N4_C100}
\end{figure}

In Fig.~\ref{fig:full_H_numerics_N4_C100}, we perform numerical simulations for $(\Delta  \beta)^2_{\rm GHZ}$, where the state is realized in $P=1$ protocol step, for $N= 4$, $C = 100$ and $\gamma/\kappa = 0.1$. In simulating the state preparation protocol with the full Hamiltonian, we use the optimal parameters - the geometric phase $\phi_1$ and cavity pulse detuning $\delta_1$ obtained by the state preparation protocol at a finite $gT$ (which uses the analytic solution of the effective Hamiltonian in Eq.~\eqref{eq:H_eff} ). The cavity drive pulse $\eta(t)$ in the full Hamiltonian is obtained by inverting the pulse $\zeta(t)$ in Eq.~\eqref{eq:H_eff} with a finite value of $\Delta \gg g$ which is set by a choice of $\mathrm{max}_{t}|\eta(t)| \gg g$: $\eta(t)=-\frac{\Delta}{S(t)}(\dot{\zeta}(t)+(i\delta+\frac{\kappa}{2})\zeta(t))-\frac{2\Delta g^2}{S(t)^3}\zeta(t)\frac{d}{dt}|\zeta(t)|^2$, where $S(t)=g^2\sqrt{1-4|\zeta(t)|^2/g^2}$. Here, we find an optimal value of $\mathrm{max}_{t}|\eta(t)|$ within the bounds $(20g, 50g)$ which minimizes $(\Delta  \beta)^2_{\rm GHZ}$. The obtained optimal values are listed in Table~\ref{tab:N4_full_sim}. 

\begin{table}
    \centering
    \begin{tabular}{|c|c|c|}
    \hline
        $gT$ & $N(\Delta \beta)^2_{\mathrm{GHZ}}$& $(\theta_{0}^{\alpha}, \theta_{0}^{\beta}, \theta_{0}^{\gamma})$\\
        &&$(\phi_1, \delta_1,\theta_{1}^{\alpha}, \theta_{1}^{\beta}, \theta_{1}^{\gamma})$, $\mathrm{max}_{t}|\eta(t)|$, $\Delta_1$\\
        \hline
        \hline
         30 & 0.482 & (1.163 , 1.566 , 0.994)\\
        &&(1.514, 1.619, 0.775, 1.577 , 1.700), 55.80, 26.12\\
        \hline
         40 & 0.451 & ( 1.177, 1.565, 0.994)\\
        &&(1.518, 1.826, 0.772, 1.577, 1.714), 40.12, 25.91\\
        \hline
        60 & 0.434 & ( 1.187, 1.565, 0.994)\\
        &&(1.521, 2.024, 0.767, 1.577, 1.724), 50.18, 44.87\\
        \hline
        80 & 0.428 & ( 1.191, 1.564, 0.994)\\
        &&(1.523, 2.119, 0.769, 1.577, 1.728 ), 46.94, 50.78\\
        \hline
        90 & 0.435 & ( 1.193, 1.564, 0.994)\\
        &&( 1.523, 2.150, 0.769,1.577 , 1.729), 38.13, 44.44\\
        \hline
    \end{tabular}
    \caption{Optimal state preparation protocol parameters $\Theta_{\opt}$ minimizing $(\Delta \beta)^2_{\mathrm{GHZ}}$ for $N=4$, $C = 100$, $\gamma/\kappa = 0.1$, and the optimal $\mathrm{max}_{t}|\eta(t)|$ and corresponding $\Delta_j$ value for full Hilbert space simulations with the full Hamiltonian with inhomogeneous coupling strength (see text).  An extra rotation along $\hat z$ direction to set $\beta_{\opt}=0$ is incorporated in $\theta^{\alpha}_{1}$\cite{extra_rotation_note}.}
    \label{tab:N4_full_sim}
\end{table}

The quantum channel of the geometric phase gate (Eq.~\ref{eq:eff_channel} of the main text) is now computed here for all qubit state components $|q_n\rangle \langle q_m|$. We start with the Hamiltonian in the transformed cavity frame given in Eq.~\eqref{eq:sm_Hamiltonian_trasnformed_cavity_frame} and perform the simulations in the Hilbert space of dimensions $(n_{\rm ph} + 1) 2^n \times (n_{\rm ph} + 1) 2^m$. Here $n_{\rm ph}$ refers to the maximum number of photons used in the simulation set such that in the decay-free limit and in the infinite detuning limit ($\Delta/g \rightarrow \infty$), the probability of the system occupying photon number states beyond $|n_{\rm ph}\rangle$ is less than a cutoff $\epsilon_{\rm ph} = 10^{-8}$. For the full Hilbert space simulations in Fig.~\ref{fig:full_H_numerics_N4_C100} for $N=4$, we have for $gT = [30, 40, 60, 80, 90]$, $n_{\rm ph} = [13, 11, 9, 8, 8]$.

We see that the $(\Delta  \beta)^2_{\rm GHZ}$ obtained by including $\gamma$ decay as local jump operators in the Lindblad formalism (circle markers) are upper bounded by the those obtained by treating $\gamma$ as a non-Hermitian contribution (cross markers). The numerical values obtained for the case with inhomogeneous $g$ coupling strength (diamond markers) indicate that the protocol remains quite robust, even when operating beyond the symmetric subspace. 

\end{document}